\documentclass[svgnames,11pt,letterpaper]{amsart}
\usepackage[letterpaper, margin=1.2in]{geometry}
\usepackage{mathpazo}  
\usepackage[english]{babel}
\usepackage{fontawesome}
\usepackage[dvipsnames]{xcolor}
\usepackage{enumerate,subcaption,overpic}  
\usepackage{url}
\usepackage{amsmath,amssymb,latexsym,mathrsfs,amssymb,amsthm}
\usepackage{bbding}
\usepackage{pifont}
\usepackage{etaremune}
\usepackage{enumitem}
\usepackage{graphicx}
\usepackage[normalem]{ulem}
\usepackage{amsthm}
\usepackage{setspace}
\let\vec=\mathbf

\newtheorem{theorem}{Theorem}[section]

\newtheorem{remark}[theorem]{Remark}

\theoremstyle{definition}

\newtheorem{rhp}{Riemann--Hilbert Problem}
\newcommand{\rhref}[1]{Riemann--Hilbert Problem~\ref{#1}}

\numberwithin{equation}{section}

\definecolor{NiceBlue}{HTML}{0069d6}
\definecolor{Teal}{rgb}{0.07,0.4,0.4}

\newcommand{\ii}{\ensuremath{\mathrm{i}}}
\newcommand{\ee}{\ensuremath{\mathrm{e}}}
\newcommand{\dd}{\ensuremath{\mathrm{d}}}

\DeclareMathOperator*{\res}{Res}
\let\Re=\undefined\DeclareMathOperator{\Re}{Re}
\let\Im=\undefined\DeclareMathOperator{\Im}{Im}

\usepackage[colorlinks=true,linkcolor=MidnightBlue,citecolor=MidnightBlue, urlcolor=MidnightBlue]{hyperref}

\title{Efficient computation of soliton gas primitive potentials}

\author{Cade Ballew}
\address{Cade Ballew: Department of Applied Mathematics, University of Washington, Seattle, WA, USA}
\email{ballew@uw.edu}

\author{Deniz Bilman}
\address{Deniz Bilman: Department of Mathematical Sciences, University of Cincinnati, Cincinnati, OH, USA}
\email{bilman@uc.edu}
\author{Thomas Trogdon}
\address{Thomas Trogdon: Department of Applied Mathematics, University of Washington, Seattle, WA, USA}
\email{trogdon@uw.edu}
\date{\today}

\begin{document}
\begin{abstract}
We consider the problem of computing a class of soliton gas primitive potentials for the Korteweg-de Vries equation that arise from the accumulation of solitons on an infinite interval in the physical domain, extending to $-\infty$.  This accumulation results in an associated Riemann--Hilbert problem on a number of disjoint intervals.  In the case where the jump matrices have specific square-root behavior, we describe an efficient and accurate numerical method to solve this Riemann--Hilbert problem and extract the potential.  The keys to the method are, first, the deformation of the Riemann--Hilbert problem, making numerical use of the so-called $g$-function, and, second, the incorporation of endpoint singularities into the chosen basis to discretize and solve the associated singular integral equation.
\end{abstract}




\maketitle

\section{Introduction}
A soliton gas can be thought of as a very large number of solitons, with possibly random amplitudes and phases, interacting weakly (rarefied gas) or strongly (dense gas) in a medium modeled by an integrable nonlinear wave model. 
This concept goes back to the work of Zakharov \cite{Zakharov71}. 
There has been a growing interest in the study of such solutions at the analytical, numerical, and experimental fronts in the past decade or so, and
the literature on soliton gasses is vast.
We refer the reader to the comprehensive review article \cite{SuretX24} on the subject and the references therein.
The notion behind the construction of such solutions from an integrable systems point of view is the so-called ``dressing'' method \cite{ZakharovM85}. This idea was employed in \cite{DyachenkoZZ16} to construct the so-called primitive potentials, which can be taken as initial data for a solution modeling a soliton gas. 

\begin{figure}[t]
    \centering
    \includegraphics[width=\linewidth]{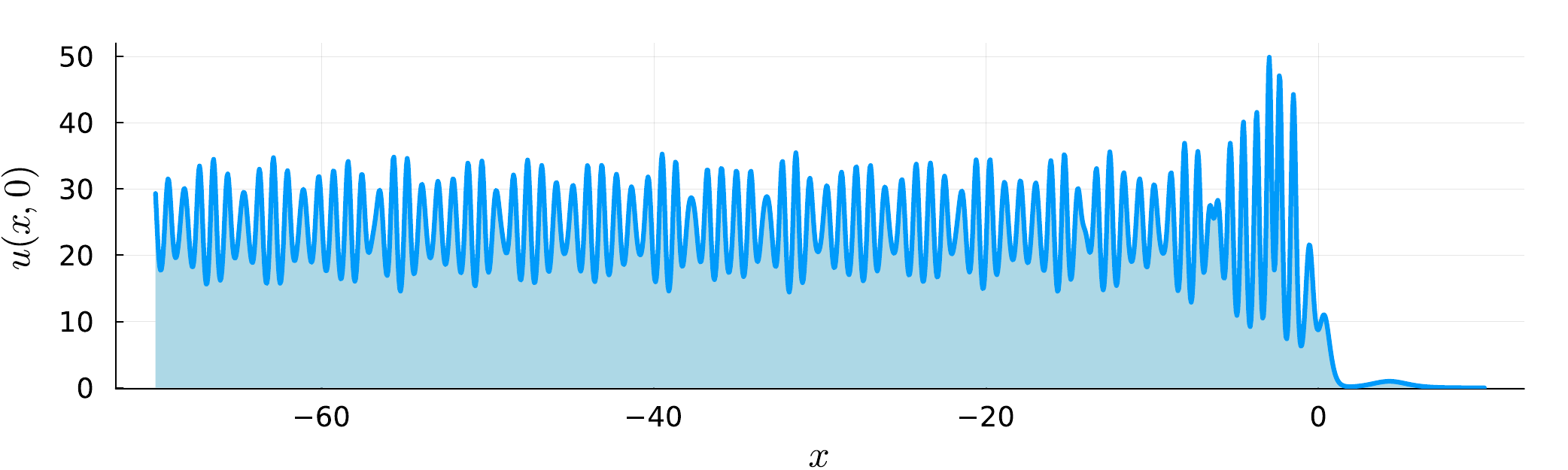}\\
    \includegraphics[width=\linewidth]{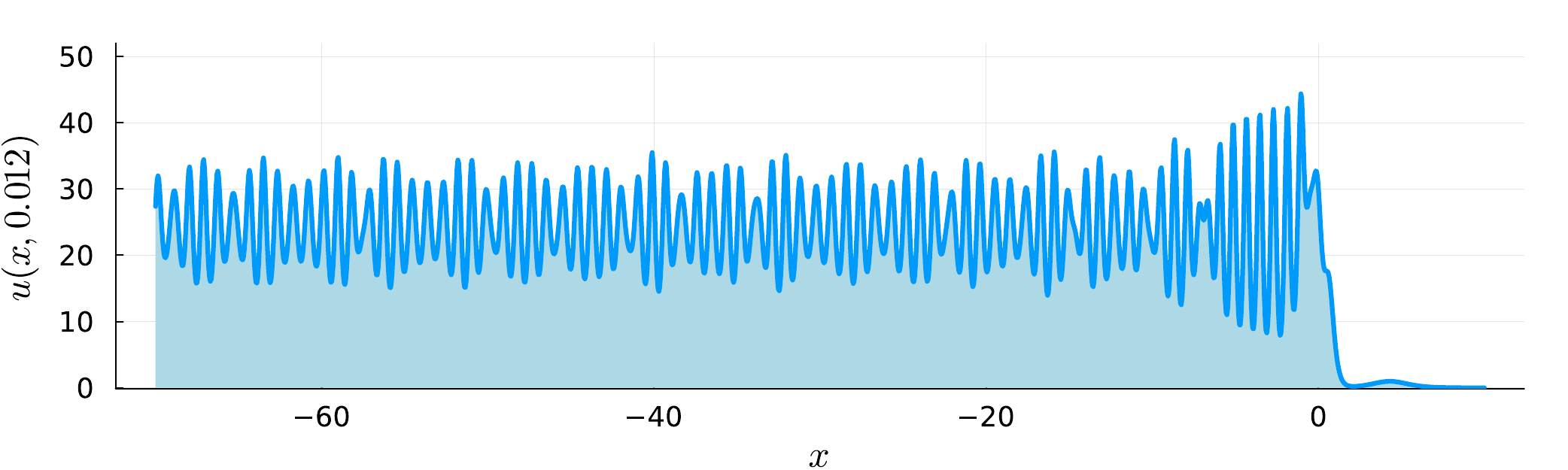}
    \caption{KdV soliton gas with $r_1(\lambda)$ supported on five pairs of bands, nonlinearly superimposed with five solitons.
    In the notation of \eqref{r1-practice}, $I_1 = (0.25,0.5)$, $I_2 = (0.8,1.2)$, $I_3=(1.5,2)$, $I_4=(2.5,3)$, $I_5=(4,5)$, with $f_1(z)=1$, $f_2(z)=1/2$, $f_3(z)=1/4$, $f_4(z)=1/8$, $f_5(z)=1/16$ and $\alpha_j=\beta_j=\frac{1}{2}$ for $j=1,\ldots,5$. The solitons are associated with (see \rhref{rhp:gas-soliton}) $\kappa_1=0.1$, $\kappa_2 = 0.7$, $\kappa_3=2.25$, $\kappa_4=3.5$, $\kappa_5 = 5.5$ with the norming constants $\chi_1=10^5$, $\chi_2=1000$, $\chi_3=100$, $\chi_4=10$, and $\chi_5=10^{-6}$.}
    \label{fig:soliton-gas-wide-1}
\end{figure}

The current state of the art for computing soliton gas solutions of integrable nonlinear wave models is largely limited to either computing a suitable nonlinear superposition of $N$ solitons for $N$ large or numerically implementing an available asymptotic formula in a suitable asymptotic region of the space-time domain. 
The former approach is equivalent to computing an iterated (or $N$-fold) Darboux transformation, and it requires high-precision arithmetic because the size of the underlying linear algebra system is proportional to $N$ (the number of solitons), and that linear system becomes numerically ill-conditioned as $N$ grows. 
The latter approach has its difficulties as the accurate evaluation of asymptotic formul\ae{} involving Riemann theta functions is a challenging task, if at all possible, especially when the genus of the underlying Riemann surface is not small.
Moreover, the availability of spacetime asymptotic formul\ae{} is limited to soliton gasses obtained by the accumulation of eigenvalues on one pair of bands in the spectral plane. Regardless, such formul\ae{} cease to be accurate for intermediate values of $(x,t)$ outside of asymptotic regimes.
An approach that is more closely related to our work was taken in \cite{DyachenkoZZ16}, where soliton gas (primitive) potentials were computed via the solution of a system of singular integral equations arising from a dressing construction.  The numerical solution of this system necessitated the use of high-precision (i.e., quadruple or higher) arithmetic. In the current work, we avoid the use of high-precision arithmetic by using Riemann--Hilbert (RH) steepest-descent deformations \cite{DeiftZ93} to, in effect, precondition the singular integral equations.
Despite the growing body of work on both theoretical and experimental fronts, an efficient framework for computing soliton gas solutions of integrable nonlinear wave models has been elusive. 

Specifically, with this article, we introduce a fast and accurate method to compute a class of soliton gas primitive potentials in the context of the Korteweg--de Vries (KdV) equation 
\begin{equation}
u_t + 6 u u_x + u_{xxx} = 0,\quad -\infty<x<+\infty,
\label{KdV}
\end{equation}
based on the numerical solution of their Riemann--Hilbert problem (RHP) representations. An advantage of this approach is that there is no time-stepping involved, as $(x,t)$ enter the problem as (explicit) parameters. The method is easily parallelizable over values of $(x,t)$.

The method presented here can compute primitive potentials of soliton gasses (at $t=0$, for all $x$) and their time evolution under the KdV flow for small values of $t$, still on the entire $x$-axis. In this regime, the method is asymptotically accurate as $x\to\pm \infty$ and does not require high-precision arithmetic. 
Importantly, it allows us to compute soliton gas solutions of the KdV equation with associated Riemann--Hilbert jump conditions supported on many pairs of bands.  We also compute such soliton gas solutions nonlinearly superposed with a number of solitons.
See Figure~\ref{fig:soliton-gas-wide-1} for a soliton gas solution of \eqref{KdV} with associated Riemann--Hilbert jump conditions  supported on five (pairs of) bands and nonlinearly superimposed with five solitons, computed with our method. Pointwise evaluation of the solution in Figure~\ref{fig:soliton-gas-wide-1} does not require higher precision arithmetic and takes about 0.7 seconds in the unmodulated region (see Section~\ref{s:unmodulated}) and 0.08 seconds in the quiescent region (see Section~\ref{s:quiet}) on a standard laptop\footnote{All computations in this paper can be performed on a Lenovo laptop running Ubuntu version 20.04 with 8 cores and 16 GB of RAM with an Intel\textregistered{} Core\texttrademark{} i7-11800H processor running at 2.30 GHz. However, due to the ease of parallelization, a cluster can greatly reduce the time it takes to generate the figures.}. Solution animations and \texttt{Julia} code used to generate the plots in this paper can be found at \cite{coderepo}.

\begin{figure}
    \centering
    \includegraphics[width=\linewidth]{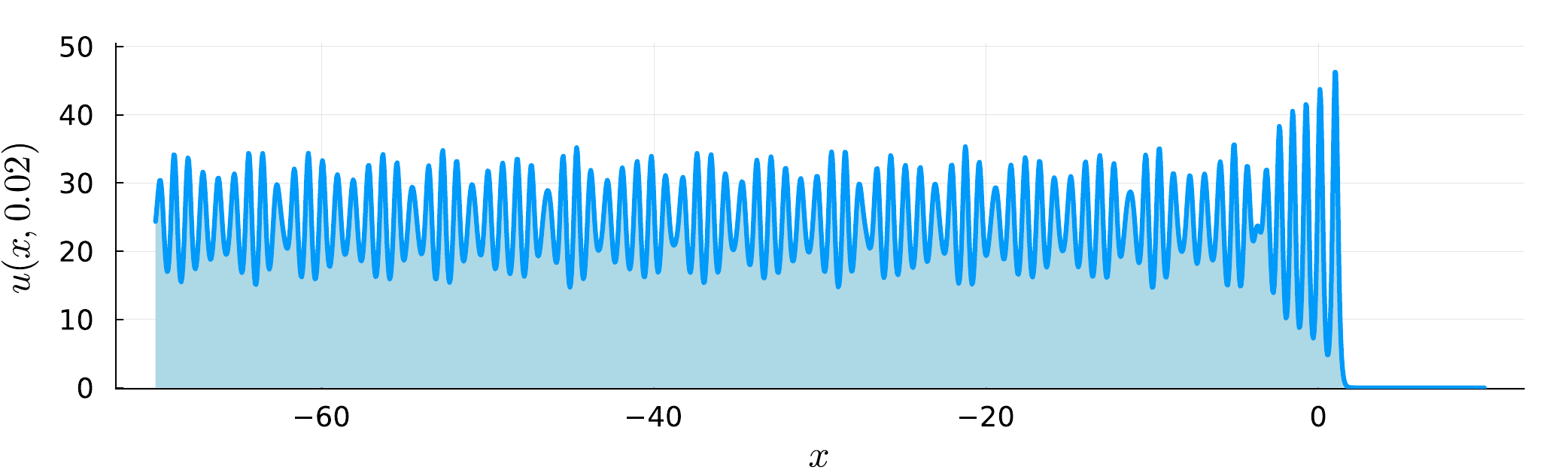}\\
    \includegraphics[width=\linewidth]{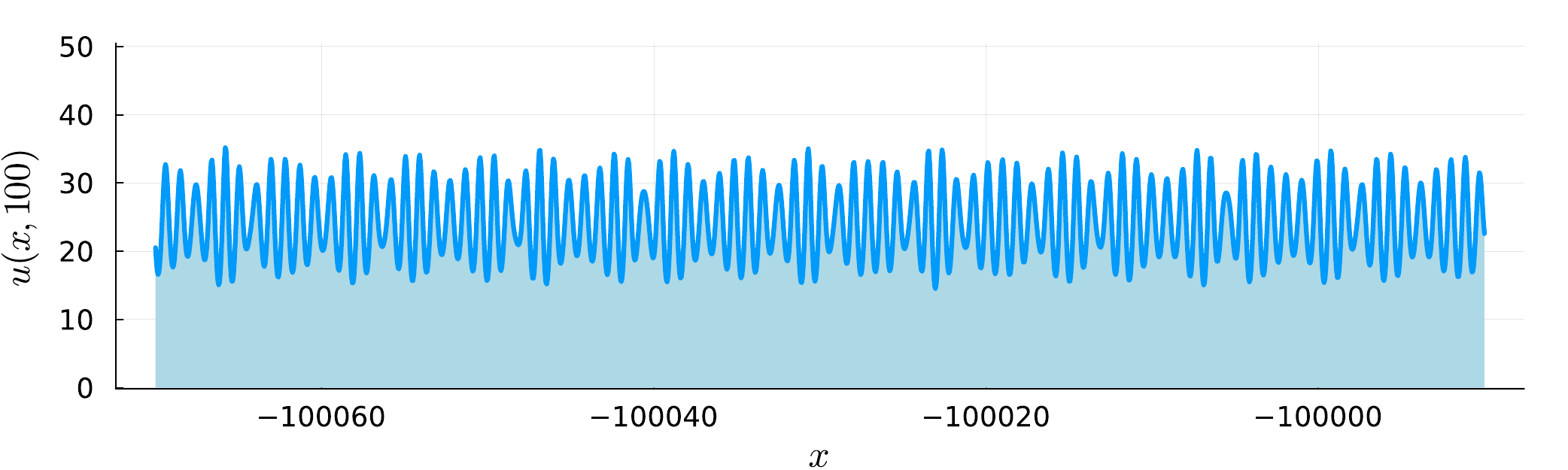}
    \caption{A pure KdV soliton gas with $r_1(\lambda)$ supported on five pairs of bands. In the notation of \eqref{r1-practice}, $I_1=(0.25,0.5)$, $I_2=(0.8,1.2)$, $I_3=(1.5,2)$, $I_4=(2.5,3)$, $I_5=(4,5)$ with $f_1(z)=1$, $f_2(z)=1/2$, $f_3(z)=1/4$, $f_4(z)=1/8$, $f_5(z)=1/16$ and $\alpha_j=\beta_j=\frac{1}{2}$ for $j=1,\ldots,5$.}
    \label{fig:soliton-gas-tail}
\end{figure}

\begin{figure}
\includegraphics[width=\linewidth]{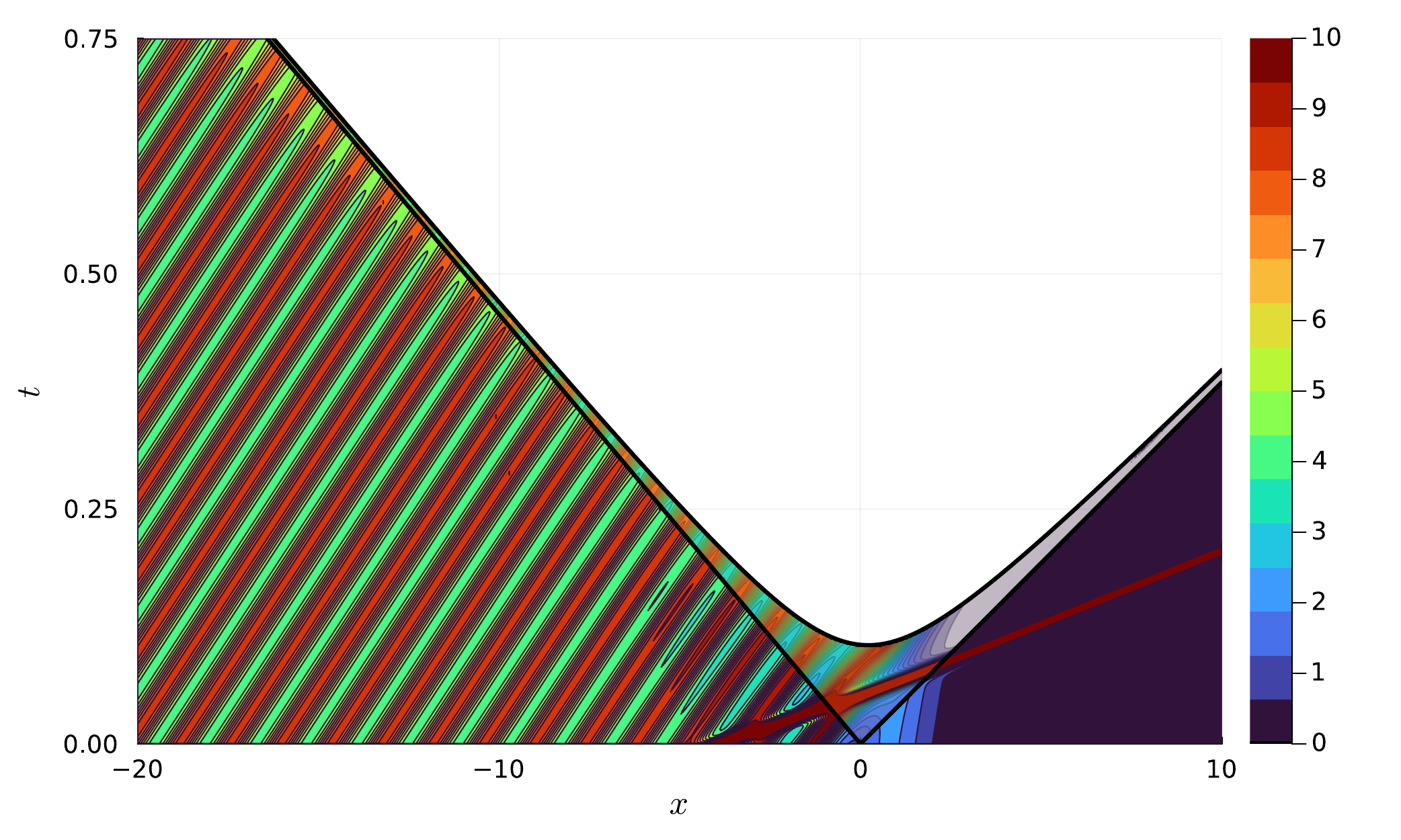}
    \caption{Density plot of the computed soliton gas with $r_1(\lambda)$ supported on a single pair of bands with two solitons. In the notation of \eqref{r1-practice}, $I_1=(1.5,2.5)$, $f_1(z)=1$, and $\alpha_1=\beta_1=\frac{1}{2}$. The solitons are associated with the eigenvalue parameters $\kappa_1=1$, $\kappa_2=4$ and the norming constants $\chi_1=10$, $\chi_2=10^{-10}$. Outside of the wedge region, the numerical method presented here is seen to be uniformly accurate with a computational cost that is independent of $(x,t)$.  Inside the wedge, the numerical method begins to break down, and additional RH deformations will need to be incorporated.}
\label{fig:wedge}
\end{figure}

The method presented here can also compute soliton gas solutions of the KdV equation in the entire $(x,t)$-plane outside of an unbounded wedge-shaped region emanating from $(x,t)=(0,0)$. See Figure~\ref{fig:soliton-gas-tail} for a pure soliton gas (no solitons) supported on five pairs of bands. The bottom panel presents the computed large-time evolution in the tail $x<-K t$ for some suitable choice of constant $K>0$.
Figure~\ref{fig:wedge} provides the density plot of a soliton gas with associated Riemann--Hilbert jump conditions supported on a single pair of bands, nonlinearly superimposed with two solitons, in the complement of the aforementioned wedge-shaped region. As is apparent from this figure, the validity of our method extends a little bit into the wedge (see the boundary curves in black). The extension stops once the exponential factors supported on the suitable deformed jump contours become too large as $(x,t)$ penetrates into the wedge.

\begin{figure}
\centering
\includegraphics[width=\linewidth]{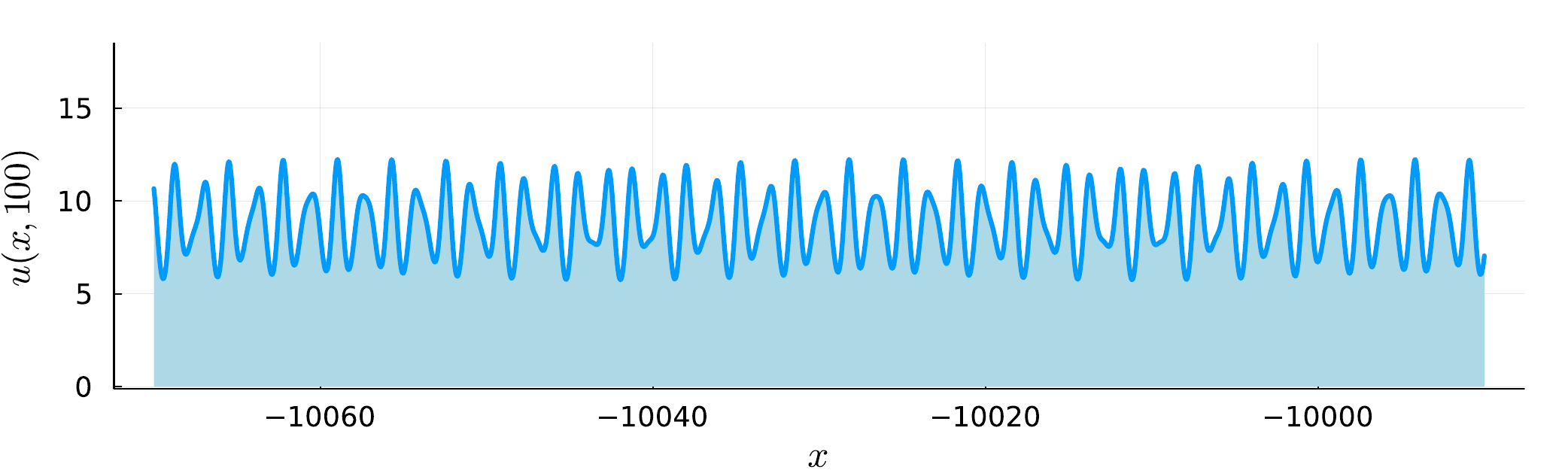}\\[2em]
\includegraphics[width=\linewidth]{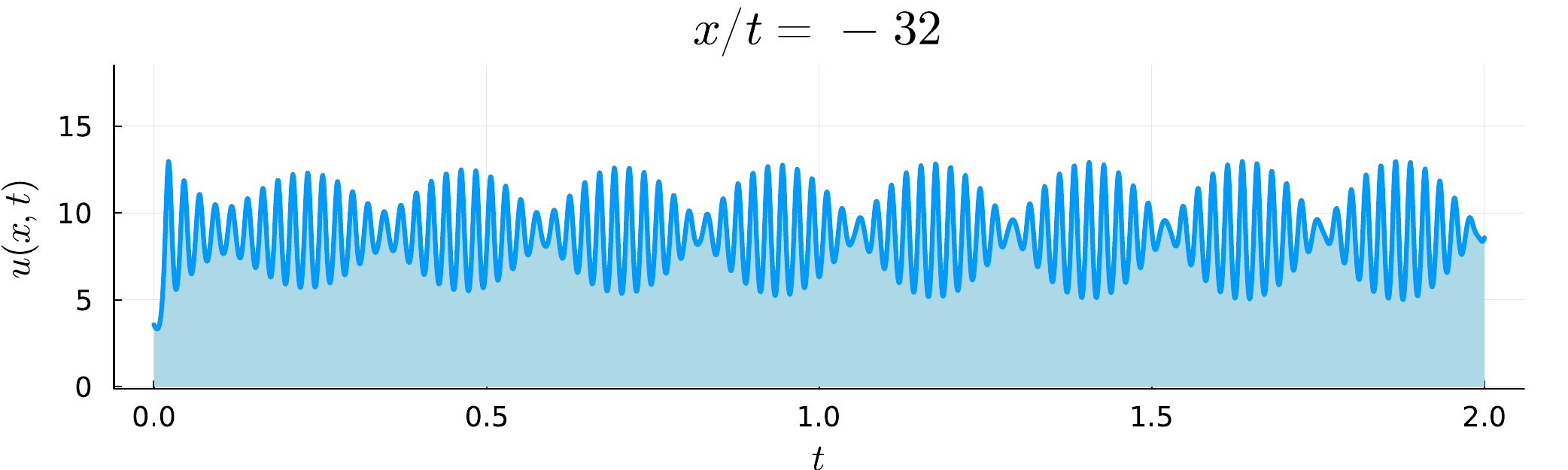}
\caption{A pure KdV soliton gas with $r_1(\lambda)$ supported on two pairs of bands. In the notation of \eqref{r1-practice}, $I_1=(1,2)$, $I_2=(2.5,3)$ with $f_1(z)=100$, $f_2(z)=1$ and $\alpha_{j}=\beta_{j}=\frac{1}{2}$, $j=1,2$. Bottom panel: The same solution plotted along the ray $x/t = -32$.}
    \label{fig:2-band}
\end{figure}

In Figure~\ref{fig:2-band}, we present a plot of the computed large-time evolution of a soliton gas with associated Riemann--Hilbert jump conditions supported on two pairs of bands (instead of five as in Figures~\ref{fig:soliton-gas-wide-1}~and~\ref{fig:soliton-gas-tail}), and in the notation of \eqref{r1-practice}, we choose $f_1(z)$ to be much larger than $f_2(z)$. 

This work is a culmination of recent advances made in computing orthogonal polynomials that are orthogonal on multiple disjoint intervals \cite{BallewT24} and work in computing large-genus solutions of the KdV equation \cite{BilmanNT22}.  The key ingredient behind these developments is the choice of a basis that encodes the behavior (e.g., the singularity structure) of the solution of an RHP at the endpoints of the jump contour. 
We also make use of the machinery available in the {\tt OperatorApproximation.jl} software package \cite{OperatorApprox}.

This work focuses mainly on the computation of soliton gas potentials supported on many bands and completing the groundwork for nonlinear superpositions of such potentials with many solitons. 
Extensions of our method to the entire $(x,t)$-plane, along with routines to compute a broader class of soliton gas solutions, will appear in a forthcoming article.

\subsection*{Acknowledgements}
This work was supported by the National Science Foundation under Grant No.\@ DMS-2108029 (DB) and Grant No.\@ DMS-2306438 (TT). Any opinions, findings, and conclusions or recommendations expressed in this material are those of the authors and do not necessarily reflect the views of the National Science Foundation. Parts of
this work were completed on Hyak, UW’s high performance computing cluster.

\subsection{Riemann--Hilbert definition of a KdV soliton gas}
Let $n\in\mathbb{Z}_+$ be fixed and consider a collection of disjoint intervals $\ii(a_j, b_j)$, $1\leq j \leq n$, on the imaginary axis with $0< a_j < b_j<a_{j+1}$ for each $1\leq j \leq n-1$. We denote the union of the these intervals and of their reflections by
\begin{equation}
\Sigma_+ := \bigcup_{j=1}^n (a_j, b_j)\qquad \text{and} \qquad \Sigma_- := \bigcup_{j=1}^n (-b_j, -a_j),
\end{equation}
and take all of these intervals to be oriented \emph{upwards}. 
We consider a function $r_1 \colon (\ii\Sigma_+ \cup \ii \Sigma_-) \to \mathbb{C}$ such that $r_1(\lambda)$ is positive for $\lambda$ on each interval $\ii (a_j, b_j)$, extending analytically to a neighborhood of each $\ii (a_j, b_j)$ with the local behavior
\begin{equation}
r_1(\ii z) = f_{a_j}(z) (z-a_j)^{\alpha_j},\quad \alpha_j\in\left\{-\tfrac{1}{2},\tfrac{1}{2}\right\},
\label{endpt-a}
\end{equation}
for $z$ in a neighborhood of $a_j$, where $f_{a_j}(z)$ is analytic and satisfies $f_{a_j}(z)>0$ for $z\geq a_j$. Similarly,
\begin{equation}
r_1(\ii z) = f_{b_j}(z) (b_j-z)^{\beta_j},\quad \beta_j\in\left\{-\tfrac{1}{2},\tfrac{1}{2}\right\},
\label{endpt-b}
\end{equation}
for $z$ in a neighborhood of $b_j$, where $f_{b_j}(z)$ is analytic and satisfies $f_{b_j}(z)>0$ for $z\leq b_j$. 
Here, all of the roots are taken to be the principal branch. 
Therefore, in the computations performed for this paper, we take
\begin{equation}
r_1(\ii z) = f_j(z)(z-a_j)^{\alpha_j}(b_j-z)^{\beta_j},\quad \text{for}~z\in I_j:=(a_j,b_j), \quad \Sigma_+=\bigcup_{j=1}^n I_j,
\label{r1-practice}
\end{equation}
for analytic functions $f_j$ and constants $\alpha_j,\beta_j\in\left\{-\tfrac{1}{2},\tfrac{1}{2}\right\}$. For $\lambda\in\ii\Sigma_-$, $r_1$ is defined by symmetry: $r_1(\lambda)=r_1(-\lambda)$. With these ingredients, we consider the following problem, which we take as the definition of a soliton gas:

\begin{rhp}[Pure KdV soliton gas]
Let $(x,t)\in\mathbb{R}\times\mathbb{R}_{+}$ be fixed. Find a $1\times 2$ row vector-valued function $\mathbf{M}(\lambda)\equiv \mathbf{M}(\lambda;x,t)$ with the following properties:
\begin{itemize}
\item[]\textbf{Analyticity:} $\mathbf{M}(\lambda)$ is analytic for $\lambda\in\mathbb{C}\setminus\mathrm{cl}( \ii\Sigma_+ \cup \ii \Sigma_-)$, admitting continuous boundary values on\footnote{$\mathrm{cl}(\Sigma)$ denotes the closure of a set $\Sigma$.} $\ii\Sigma_+ \cup \ii \Sigma_-$.
\item[]\textbf{Jump conditions:} The boundary values $\mathbf{M}^+(\lambda)$ (resp.\@ $\mathbf{M}^-(\lambda)$) from the left (resp.\@ right) are related via
\begin{align}
\mathbf{M}^+(\lambda)&=\mathbf{M}^-(\lambda)\begin{bmatrix} 1 & 0 \\ -2\ii r_1(\lambda)\ee^{2\ii(x \lambda + 4 t \lambda^3)} & 1\end{bmatrix},\quad \lambda\in \ii(a_j,b_j),\\
\mathbf{M}^+(\lambda)&=\mathbf{M}^-(\lambda)\begin{bmatrix} 1 &  2\ii r_1(\lambda)\ee^{-2\ii(x\lambda + 4 t \lambda^3)} \\ 0 & 1\end{bmatrix},\quad \lambda\in \ii(-b_j,-a_j), \quad j=1,2,\ldots,n.
\end{align}
\item[]\textbf{Normalization:} $\mathbf{M}(\lambda)=\begin{bmatrix} 1 & 1\end{bmatrix} + \mathrm{O}(\lambda^{-1})$ as $\lambda\to\infty$.
\item[]\textbf{Symmetry condition:} $\mathbf{M}(-\lambda)=\mathbf{M}(\lambda) \begin{bmatrix} 0 & 1 \\ 1 & 0 \end{bmatrix}$.
\end{itemize}
\label{rhp:gas}
\end{rhp}
This problem has appeared in \cite{DyachenkoZZ16, GirottiX21} with different assumptions on the behavior of $r_1(\lambda)$ at the endpoints\footnote{A derivation of \rhref{rhp:gas} via a limiting procedure involving the accumulation of eigenvalues of the Schr\"odinger operator on a single pair of bands with respect to a suitable density can be found in \cite{GirottiX21}.}. The setting of \cite{GirottiX21} assumes that $r_1(\lambda)$ is nonvanishing and bounded at the endpoints, in contrast with the assumptions in our work. As the reader will see, the endpoint behavior that we assume for $r_1(\lambda)$ enables a particularly fast computational method.
\rhref{rhp:gas} has a unique solution \cite{GirottiX21} (see Section~\ref{s:local-solve} for the notion of a solution when $\alpha_j$ or $\beta_j$ is negative), and we can recover the soliton gas solution of \eqref{KdV} by the second-order term at infinity \cite{EgorovaPT24}
\begin{equation}
u(x,t) = -\lim_{\lambda\to\infty} 2\lambda^2(m_1(z;x,t) m_2(z;x,t) - 1),
\label{recovery-X}
\end{equation}
where $m_j(z;x,t)$ is the $j$th component of the row vector $\mathbf{M}(\lambda;x,t)$.

We proceed with rotating the $\lambda$-plane clockwise by introducing $r(z) :=2 r_1(\ii z)$ and $\mathbf{Y}(z):=\mathbf{M}(\ii z)$, which satisfies the jump conditions
\begin{alignat}{2}
\mathbf{Y}^+(z)&=\mathbf{Y}^-(z) \begin{bmatrix} 1 & 0 \\ -\ii r(z)\ee^{-2\theta(z;x,t)} & 1\end{bmatrix},&&\quad z\in \Sigma_+,\qquad \theta(z;x,t):= xz - 4 tz^3,\label{Y-jump-plus}\\
\mathbf{Y}^+(z)&=\mathbf{Y}^-(z)\begin{bmatrix} 1 &  \ii r(z)\ee^{2\theta(z;x,t)} \\ 0 & 1\end{bmatrix},&&\quad z\in \Sigma_-\label{Y-jump-minus},
\end{alignat}
and is analytic elsewhere. Note that all (real) intervals are oriented in the increasing direction. We consider the numerical solution of the RHP satisfied by $\mathbf{Y}(z)$, which inherits the normalization and symmetry conditions in \rhref{rhp:gas}.

Regarding the behavior of $\mathbf{Y}(z)$ at the endpoints,
let $c_j$ denote either of the endpoints $a_j$ or $b_j$, and let $\gamma_j$ denote the corresponding power $\alpha_j$ or $\beta_j$ as in \eqref{endpt-a} and \eqref{endpt-b}. 
As $z\to c_j$, we have
\begin{equation}
\mathbf{Y}(z) = 
\begin{cases} 
\begin{bmatrix} \mathrm{O}(1) & \mathrm{O}(1) \end{bmatrix}, &\quad \text{if} \quad \gamma_j = \frac{1}{2},\\[0.5em]
\begin{bmatrix} \mathrm{O}(|z-c_j|^{-\frac{1}{2}}) & \mathrm{O}(1) \end{bmatrix}, &\quad \text{if} \quad\gamma_j = -\frac{1}{2}.
\end{cases}
\label{Y-endpt}
\end{equation}
This situation is mirrored on the negative real axis. As $z\to-c_j$, we have
\begin{equation}
\mathbf{Y}(z) = 
\begin{cases} 
\begin{bmatrix} \mathrm{O}(1) & \mathrm{O}(1) \end{bmatrix}, &\quad \text{if} \quad \gamma_j = \frac{1}{2},\\[0.5em]
\begin{bmatrix}  \mathrm{O}(1) &\mathrm{O}(|z+c_j|^{-\frac{1}{2}})\end{bmatrix}, &\quad \text{if} \quad\gamma_j = -\frac{1}{2}.
\end{cases}
\label{Y-endpt2}
\end{equation}
Therefore, there is no ambiguity in the statement of the RHP for $\mathbf{Y}(z)$ (or in \rhref{rhp:gas}, for that matter) arising from unspecified behavior of the solution at the endpoints of the jump contour --- it is implied by the behavior of the jump matrix at the endpoints.

Writing $\lambda = \ii z$ in the large-$\lambda$ expansions that led to \eqref{recovery-X}, the soliton gas solution $u(x,t)$ of \eqref{KdV} is recovered from $\mathbf{Y}(z)\equiv \mathbf{Y}(z;x,t)$ via the formula
\begin{equation}
u(x,t) = \lim_{z\to\infty} 2 z^2 (y_1(z;x,t)y_2(z;x,t) - 1).
\label{recovery}
\end{equation}

\subsection{From local solutions to consistent numerical approximations}
\label{s:local-solve}
A typical approach to the numerical solution of Riemann--Hilbert Problem~\ref{rhp:gas}, for example, is to seek a singular integral equation that the Cauchy density solves \cite{Olver12,TrogdonO-book}.  More specifically, if
\begin{align}
    \vec Y^+(z) = \vec Y^-(z) \vec J(z), \quad z \in \Gamma,
\end{align}
then we write
\begin{align}
    \vec Y(z) = \begin{bmatrix} 1 & 1 \end{bmatrix} + \mathcal C_{\Gamma} \vec U(z), \quad \mathcal C_{\Gamma}\vec U(z) = \frac{1}{2 \pi \ii} \int_{\Gamma} \frac{\vec U(\zeta)}{\zeta - z} \dd \zeta.
\end{align}
If we denote the boundary values of the Cauchy transform by $\mathcal C_{\Gamma}^\pm$, $\vec U$ satisfies the singular integral equation
\begin{align}
    \mathcal C_{\Gamma}^+\vec U(z) - \mathcal C_{\Gamma}^- \vec U(z) \vec J(z) = \begin{bmatrix} 1 & 1 \end{bmatrix}(\vec J(z) - \mathbb{I}), \quad z \in \Gamma.
\end{align}
To construct an efficient numerical method for this equation, one must choose a basis in which $\vec U$ can be accurately represented, something that will typically result in the consistency of the numerical method.  To illustrate this, we consider a toy problem where $\Gamma = \Gamma_1 \cup \Gamma_2$ are disjoint, $\Gamma_1 = [-1,1]$, oriented from $z=-1$ to $z=1$, and
\begin{align}
    \vec J|_{\Gamma_1}(z) = \begin{bmatrix} 0 & \ii h(z) \\ \ii /h(z) & 0 \end{bmatrix}, \quad h(z) = \sqrt{1 -z^2 }.
\end{align}
We will build a local solution to this problem near $z = 1$, and it will tease out the singularity structure.  Technically speaking, we will also need to verify that our local solution satisfies the requisite endpoint conditions, see \eqref{S-endpt}, for example.  We use this example because it represents the structure encountered in the unmodulated region of Section~\ref{s:unmodulated}.  For that encountered in the quiescent region of Section~\ref{s:quiet}, the same analysis can be performed, but it is simpler because the jump matrices are triangular and local solutions are found directly by Cauchy integrals.

Consider $\ell(z) = (z-1)^{1/4} (z + 1)^{1/4}$, where we use the principal branch cut along $(-\infty,0]$ for the quarter root function.  It follows that this function satisfies $\ell^+(z) \ell^-(z) = h(z)$, $z \in [-1,1]$.  Then, define
\begin{align}
    \vec L(z) = \ell(z)^{-\sigma_3}, \quad \sigma_3 = \mathrm{diag}(1,-1).
\end{align}
We compute, in a neighborhood of $z = 1$,
\begin{align}
         \vec L^-(z)\vec J|_{\Gamma_1}(z)     \vec L^+(z)^{-1} = \begin{bmatrix} 0 & \ii \frac{h(z)}{ \ell^+(z) \ell^-(z)}  \\ \ii  \frac{ \ell^+(z) \ell^-(z)}{h(z)} & 0 \end{bmatrix} = \begin{bmatrix} 0 & \ii \\ \ii & 0 \end{bmatrix}.
\end{align}
Then, we factorize
\begin{align}
    \begin{bmatrix} 0 & \ii \\ \ii & 0 \end{bmatrix} = \vec Q  a^-(z)^{-\sigma_3} a^+(z)^{\sigma_3} \vec Q^{-1}, \quad \vec Q = \frac{1}{\sqrt{2}} \begin{bmatrix} 1 & -1 \\ 1 & 1 \end{bmatrix}, \quad a(z) = \left( \frac{z -1}{z+ 1}\right)^{1/4},
\end{align}
giving
\begin{align}
    a^-(z)^{\sigma_3}\vec Q^{-1} \vec L^-(z)\vec J|_{\Gamma_1}(z)     \vec L^+(z)^{-1}\vec Q a^+(z)^{-\sigma_3} = \mathbb{I}.
\end{align}
All of this is to say that
\begin{align}
    \boldsymbol{\Psi}(z) = a(z)^{\sigma_3} \vec Q^{-1} \vec L(z) = \frac{1}{\sqrt{2}}\begin{bmatrix}
 \frac{1}{\sqrt{z+1}} & \sqrt{z-1} \\
 -\frac{1}{\sqrt{z-1}} & \sqrt{z+1} \\
 \end{bmatrix},
\end{align}
is a local solution to the RHP near $ z = 1$, where the square roots denote the principal branch cut along $(-\infty,0]$. Then define, for $\epsilon$ small,
\begin{align}
    \widehat{\vec Y}(z) = \begin{cases} \vec Y(z) & |z-1|> \epsilon,\\
    \vec Y(z) \boldsymbol{\Psi(z)}^{-1} & |z-1|< \epsilon,\end{cases}
\end{align}
which will be an analytic function near $z = 1$ (using the requisite endpoint conditions).  We find the representation
\begin{align}
    \vec Y(z) = \widehat{\vec Y}(z) \boldsymbol{\Psi}(z),
\end{align}
for $z$ near $z = 1$.  To back out the behavior of $\vec U$, we use the Plemelj lemma,
\begin{align}
    \vec U(z) =  \vec Y^+(z) -  \vec Y^-(z) = \widehat{\vec Y}(z) \left(\boldsymbol{\Psi}^+(z) - \boldsymbol{\Psi}^-(z)\right) = \widehat{\vec Y}(z) \begin{bmatrix} 0& \sqrt{2(z-1)} \\ -\sqrt{\frac{2}{z-1}} & 0 \end{bmatrix}, \quad |z-1| < \epsilon.
\end{align}
A similar analysis can be performed near $z = -1$, and patching together these arguments, it follows that
\begin{align}
    \vec U|_{\Gamma_1}(z) = \vec A(z)\begin{bmatrix}  0& \sqrt{1-z^2}\\ \frac{1}{\sqrt{1-z^2}} &0\end{bmatrix},
\end{align}
where $\vec A(z)$ is an analytic vector-valued function that can therefore be approximated well with polynomials of low degree.  This local analysis informs the ansatz \eqref{ansatz} below.

\section{Computing soliton gasses}
\subsection{Quiescent region}
\label{s:quiet}
When $t=0$, the jump matrices in \eqref{Y-jump-plus}--\eqref{Y-jump-minus} are $\mathrm{O}(1)$ (away from the endpoints) for $x\geq 0$ and become exponentially close to identity as $x\to+\infty$ since $\Re(-2 z x) < 0$ on $\Sigma_+$ and  $\Re(2 z x) < 0$ on $\Sigma_-$ in this case. 
This configuration extends to a region of the $(x,t)$-plane for $t>0$ as long as $\Re(-2\theta(z;x,t))<0$ is maintained on $\Sigma_+$ (the situation is mirrored on $\Sigma_-$ automatically). The axis $\Re(z)=0$ is always a part of the locus $\Re(-2\theta(z;x,t))=0$. The other branch of this locus is given by $\Re(z)^2 = 3 \Im(z)^2 + \frac{x}{4t}$. 
Thus, the aforementioned boundedness or exponential decay of the jump matrices to the identity is preserved for all $(x,t)$ with $t \geq 0$ satisfying
\begin{equation}
x \geq 4 K t,
\label{quiet-region}
\end{equation}
for an arbitrary constant $K>b_n^2$.
In this setting, the RHP at hand can be treated numerically as is, without implementing any contour deformations, by using an appropriate basis with weights encoding the endpoint behavior of $\mathbf{Y}(z)$. 
We employ the approach put forth in \cite{BallewT24,BilmanNT22} and consider the ansatz given by the superposition of weighted Cauchy transforms
\begin{equation}
\mathbf{Y}(z) = \begin{bmatrix} 1 & 1 \end{bmatrix} + \sum_{j=1}^n\left[ \int_{a_j}^{b_j} \frac{\mathbf{F}_{+j} (\zeta) {\begin{bmatrix}w_{+j}^{[1]}(\zeta)& 0 \\ 0 & w_{+j}^{[2]}(\zeta)\end{bmatrix}}\dd \zeta}{\zeta-z} +  \int_{-b_j}^{-a_j} \frac{\mathbf{F}_{-j} (\zeta) { \begin{bmatrix}w_{-j}^{[1]}(\zeta)& 0 \\ 0 & w_{-j}^{[2]}(\zeta)\end{bmatrix}}\dd \zeta}{\zeta-z}\right],
\label{ansatz}
\end{equation}
for unknown row vector-valued densities $\mathbf{F}_{\pm j}(\zeta)$ on each interval. Here, each of the weights $w_{\pm j}^{[1]}(\zeta)$ and $w_{\pm j}^{[2]}(\zeta)$ (used for the first and second columns, respectively) is equal to one of the Chebyshev weights
\begin{align*}
\rho_1(\zeta)&:= \frac{1}{\sqrt{(\zeta-a)(b-\zeta)}}, 
\quad \rho_2(\zeta):=\rho_1(\zeta)^{-1}, \\
\rho_3(\zeta)&:=\sqrt{(\zeta-a)(b-\zeta)^{-1}}, 
\quad \rho_4(\zeta):=\rho_3(\zeta)^{-1},
\end{align*}
for the four kinds of Chebyshev orthogonal polynomials on an interval $[a,b]$ (on $[a_j, b_j]$ for $w_{+j}^{[1],[2]}(\zeta)$ and on $[-b_j,-a_j]$ for $w_{-j}^{[1],[2]}(\zeta)$).
The choice is made so that the weight captures the behavior of the given $r(z)$ (and hence that of $\mathbf{Y}(z)$) at the endpoints, and the corresponding Chebyshev polynomials are used as the basis on the relevant interval. 
The associated singular integral equation is discretized and solved by collocation, as described in \cite{Olver12, TrogdonO-book}, by enforcing that the singular integral equation should be satisfied exactly at mapped Chebyshev first-kind zeros.
\begin{remark}
The triangular nature of the jump matrices \eqref{Y-jump-plus}--\eqref{Y-jump-minus} implies that the second column of $\mathbf{F}_{+j}(\zeta)$ and the first column of $\mathbf{F}_{-j}(\zeta)$ are identically equal to $0$. Therefore, the weights $w^{[2]}_{+j}(\zeta)$ and $w^{[1]}_{-j}(\zeta)$ are immaterial. In practice, we do not encode this structure and allow the linear system arising from collocation to force those entries to be equal to $0$. This structure is lost in other asymptotic regions upon contour deformations; therefore, we present the ansatz \eqref{ansatz} in its general form for future reference.
\end{remark}
Finally, note that since $r(z)=r(-z)$, $w_{-j}^{[2]}(\zeta) = w_{+j}^{[1]}(-\zeta)$ for $\zeta\in[-b_j, -a_j]$. 
The region described by \eqref{quiet-region} is the analogue of the ``constant region'' as $t\to+\infty$ in \cite{GirottiX21} for the case $n=1$.

\subsection{Unmodulated region}
\label{s:unmodulated}
\begin{figure}
\begin{center}
\includegraphics[width=.8\textwidth]{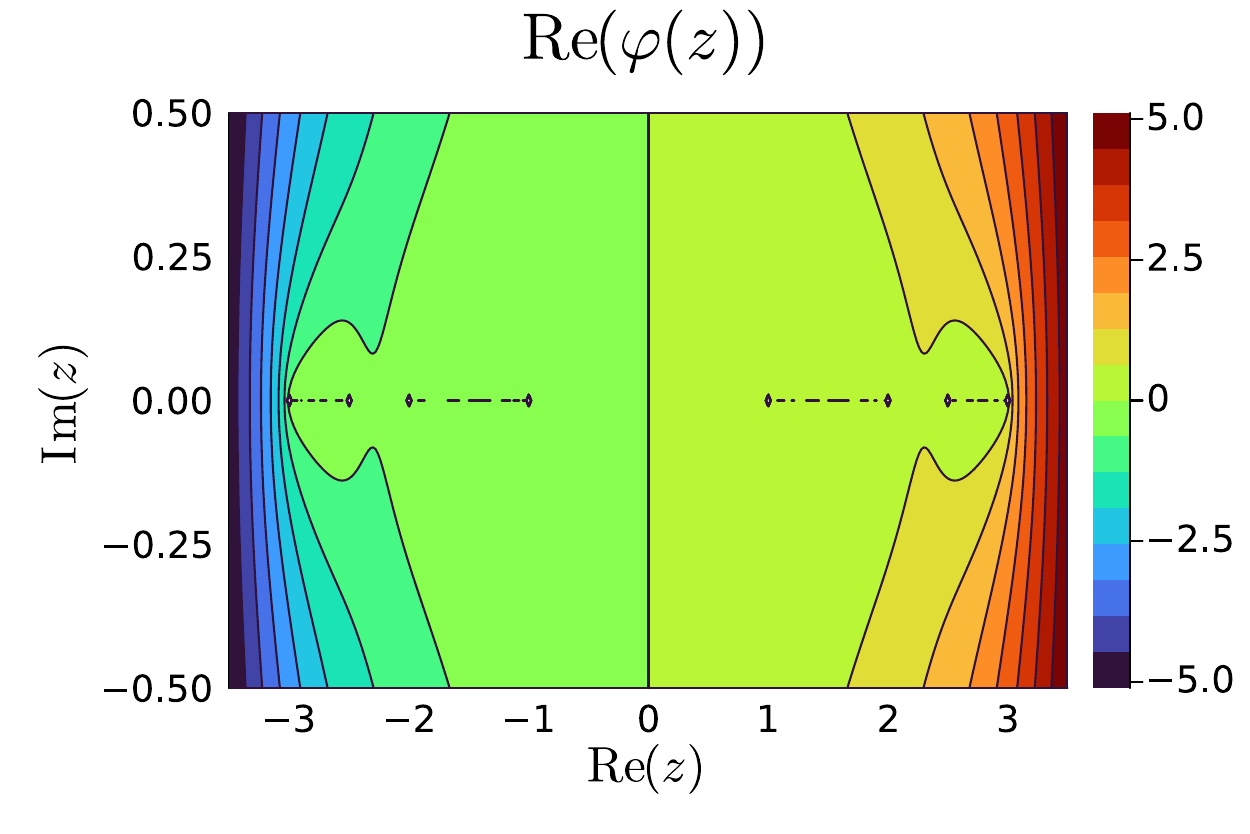}
    \caption{Contour plot showing the sign of $\mathrm{Re}(\varphi(z;x,t))$ in the complex plane for $t>0$ in the unmodulated region treated in Section~\ref{s:unmodulated}. Here, $x=-2$, $t=0.01$, and in the notation of \eqref{r1-practice}, $I_1=(1,2)$, $I_2=(2.5,3)$.}
\label{f:re-phi-sign}
\end{center}
\end{figure}
When $t=0$ and $x<0$, the situation is dramatically different: All of the exponentials in \eqref{Y-jump-plus}--\eqref{Y-jump-minus} grow exponentially and unboundedly as $x\to-\infty$. 
This behavior cannot be put under control solely by employing one of the four canonical matrix factorizations for the jump matrices since the exponents are real-valued. In this setting, the RHP is stabilized by modifying the exponent via introducing a $g$-function. This is a generalization of the approach taken in \cite{GirottiX21, GirottiX23} for the case $n=1$.
Let $\Gamma_{\pm j}$ denote the gaps: $\Gamma_0=[-a_1, a_1]$, $\Gamma_{j}=[b_{j},a_{j+1}]$, and $\Gamma_{-j}=[-a_{j+1},-b_{j}]$ for $j=1,2,\ldots,n-1$.
To have a uniform treatment that also captures a region when $t>0$, we proceed with allowing for $t\geq 0$ with $x<0$ and do not designate a particular large parameter.
We seek a function $g(z)\equiv g(z;x,t)$ analytic for $z\notin [-b_n , b_n]$, with the following additional properties: $g(z)$ admits continuous boundary values $g^{\pm}(z)$ on $[-b_n,b_n]$, taken from $\mathbb{C}^{\pm}$, satisfying
\begin{alignat}{2}
g^+(z)+g^-(z) &= 2 \theta(z;x,t), &&\quad z\in \Sigma_+ \cup \Sigma_-,\\
g^+(z)-g^-(z) &= \ii \Omega_{\pm j}(x,t), &&\quad z\in \Gamma_{\pm j},\quad j=0,1,2,\ldots,n-1,
\end{alignat}
where $\Omega_{\pm j}(x,t)$ are real-valued constants. 
Finally, $g(z)=\mathrm{O}(z^{-1})$ as $z\to\infty$.
Let $R(z)$ be the function analytic for $z\notin \mathrm{cl}(\Sigma_+ \cup \Sigma_-)$ satisfying 
$
R(z)^2 = \prod_{j=1}^n (z^2-a_j^2)(z^2-b_j^2)
$
along with $R(z)=z^{2n} + \mathrm{O}(z^{2n-2})$ as $z\to\infty$. Then,
\begin{equation}
g(z) = \frac{R(z)}{2\pi \ii} \int\limits_{\Sigma_{+} \cup \Sigma_{-}} \frac{2\theta(\zeta;x,t) \dd \zeta}{R^+(\zeta)(\zeta-z)} + \sum_{\ell=-(n-1)}^{n-1}\frac{\Omega_\ell(x,t)R(z)}{2\pi} \int\limits_{\Gamma_\ell} \frac{\dd \zeta}{R(\zeta)(\zeta - z)}\,,
\end{equation}
where the $2n-1$ constants $\Omega_{\ell}(x,t)$ are chosen to ensure the desired decay $g(z)=\mathrm{O}(z^{-1})$ as $z\to\infty$. 
These moment conditions result in a linear system:
\begin{equation}\label{g-sys}
\sum_{\ell=-(n-1)}^{n-1}\ii\Omega_{\ell}(x,t) \int\limits_{\Gamma_\ell} \frac{\zeta^k \dd \zeta}{R(\zeta)} = - \int\limits_{\Sigma_-\cup\Sigma_+} \frac{\zeta^k \theta(\zeta;x,t)}{R^+(\zeta)} \dd\zeta,\quad k =0,1,2,\ldots, 2n-2.
\end{equation}
While this linear system can be shown to have a unique solution, it becomes ill-conditioned for large values of $n$ due to the presence of the monomials $\zeta^k$. We solve an equivalent but empirically well-conditioned linear system obtained by replacing the monomials $\zeta^k$ with a suitable basis of polynomials of degree $2n-2$, chosen to vanish at the midpoints of the $2n-2$ gaps. That is, we compute the constants $\Omega_\ell(x,t)$ using the basis
\begin{equation*}
p_{k}(x)=\prod_{j=-n+1, j\neq k}^{n-1}(x-\mu_j), \quad k=-n+1,\ldots,n-1,
\end{equation*} 
where $\mu_j$ denotes the midpoint of $\Gamma_j$. We compute the integrals in the linear system \eqref{g-sys} via Gauss--Chebyshev quadrature.

\begin{figure}
\includegraphics[width=\textwidth]{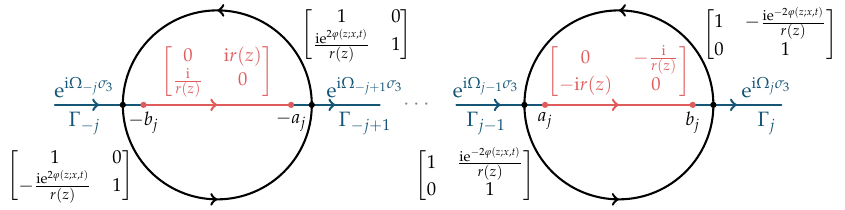}
\caption{Jump contours and jump matrices associated with $\mathbf{S}(z;x,t)$ near each interval. $\varphi(z;x,t) = g(z;x,t) - \theta(z;x,t)$.}
\label{f:S-jumps}
\end{figure}

Now, introduce open and disjoint disks $D_j$, $j=1,2,\ldots,n$, chosen so that $D_j$ encloses $[a_j,b_j]$ in its interior,
and 
let $D_{-j}$ be the analogue of $D_{j}$ for $[-b_j, -a_j]$. We take the disk boundaries to be oriented counter-clockwise and
make the following definitions (with correlated signs on each line):
\begin{alignat}{2}
\mathbf{S}(z) &:= \mathbf{Y}(z)
\begin{bmatrix}
1 & \frac{\ii}{r(z)}\ee^{2\theta(z;x,t)} \\ 0 & 1
\end{bmatrix}^{\mp 1} \ee^{g(z;x,t)\sigma_3},&&\qquad z\in D_j \cap \mathbb{C}^{\pm},\label{Y-to-S-pos}\\
\mathbf{S}(z) &:= \mathbf{Y}(z)
\begin{bmatrix}
1 &0  \\  \frac{-\ii}{r(z)}\ee^{-2\theta(z;x,t)} & 1
\end{bmatrix}^{\mp 1} \ee^{g(z;x,t)\sigma_3},&&\qquad z\in D_{-j} \cap \mathbb{C}^{\pm},\label{Y-to-S-neg}
\end{alignat}
for $j=1,2,\ldots,n$. Recall that $\sigma_3 = \mathrm{diag}(1,-1)$. We set $\mathbf{S}(z):= \mathbf{Y}(z)\ee^{g(z;x,t)\sigma_3}$ elsewhere. 
Using $r^+(z)=-r^-(z)$ on $\Sigma_+\cup\Sigma_-$, we find that the jump conditions satisfied by $\mathbf{S}(z)$ are as given in Figure~\ref{f:S-jumps}.
We denote the modified exponent by $\varphi(z;x,t):=g(z;x,t)-\theta(z;x,t)$.
A contour plot of the sign of $\mathrm{Re}(\varphi(z;x,t))$ is given in Figure~\ref{f:re-phi-sign}, and one can see that the jump matrices on the circles are exponentially close to the identity matrix if $t>0$ becomes large for $x/t$ in this region. This behavior together with the fact that the circular jump contours are detached from $\Sigma_+ \cup \Sigma_-$ (no self-intersection points with jump matrices involving $r(z)$) is what enables an efficient numerical method. Therefore, the assumptions on the behavior of $r(z)$ at the endpoints are absolutely essential for our approach.

Again, let $c_j$ denote either of the endpoints $a_j$ or $b_j$, and let $\gamma_j$ denote the corresponding power $\alpha_j$ or $\beta_j$ as in \eqref{endpt-a} and \eqref{endpt-b}. From \eqref{Y-endpt}, \eqref{Y-to-S-pos}, and \eqref{Y-to-S-neg}, we find that as $z\to c_j$,
\begin{equation}
\mathbf{S}(z) = 
\begin{cases} 
\begin{bmatrix} \mathrm{O}(1) & \mathrm{O}(|z-c_j|^{-\frac{1}{2}}) \end{bmatrix}, &\quad \text{if} \quad \gamma_j = \frac{1}{2},\\[0.5em]
\begin{bmatrix} \mathrm{O}(|z-c_j|^{-\frac{1}{2}}) & \mathrm{O}(1) \end{bmatrix}, &\quad \text{if} \quad\gamma_j = -\frac{1}{2}.
\end{cases}
\label{S-endpt}
\end{equation}
The situation is again mirrored as $z$ approaches an endpoint of $[-b_j,-a_j]$, but the behavior of the columns is flipped.

As in \cite[Section 3.2.3]{BallewT24}, we now introduce a correction to the $g$-function to eliminate the constant jump conditions on the gaps $\Gamma_{\pm j}$, $j=0,1,\ldots n-1$ at the expense of introducing constants to the jump matrices on the intervals $\Sigma_+\cup\Sigma_-$. We seek a function $h(z)\equiv h(z;x,t)$ analytic for $z\notin [-b_n,b_n]$ with the properties
\begin{alignat}{3}
h^+(z) - h^-(z) &= \log(\exp(-\ii \Omega_{\pm k}(x,t))),\quad &&z\in \Gamma_{\pm k},&&\quad k=0,1,2,\ldots, n-1,\\
h^+(z) + h^-(z) &= A_{+j}(x,t),\quad &&z\in(a_j, b_j),&&\quad j=1,2,\ldots, n,\\
h^+(z) + h^-(z) &= A_{-j}(x,t),\quad &&z\in(-b_j, -a_j),&&\quad j=1,2,\ldots, n,
\end{alignat}
along with $h(z)=\mathrm{O}(z^{-1})$ as $z\to\infty$. Then,
\begin{equation}
\begin{aligned}
h(z)=&\frac{R(z)}{2\pi\ii}
\sum_{j=1}^{n}\left[\int_{a_j}^{b_j}\frac{A_{j+}(x,t)\dd \zeta}{R^+(\zeta)(\zeta-z)}+\int_{-b_j}^{-a_j}\frac{A_{j-}(x,t)\dd \zeta}{R^+(\zeta)(\zeta-z)}\right]\\&+\frac{R(z)}{2\pi\ii}\sum_{\ell=-(n-1)}^{n-1}\int\limits_{\Gamma_\ell} \frac{\log(\exp(-\ii\Omega_\ell(x,t)))\dd \zeta}{R(\zeta)(\zeta-z)},
\end{aligned}
\end{equation}
where the constants $A_{\pm j}(x,t)$ are determined to ensure the desired decay at infinity. This results in the following linear system of $2n$ equations:
\begin{equation}
\sum_{j=1}^{n} \left[
A_{+j}(x,t)\int_{a_j}^{b_j}\frac{\zeta^k \dd \zeta}{R^+(\zeta)} + A_{-j}(x,t)\int_{-b_j}^{-a_j}\frac{\zeta^k \dd \zeta}{R^+(\zeta)} \right]
= - \sum_{\ell=-(n-1)}^{n-1}\log(\exp(-\ii\Omega_\ell(x,t)))\int\limits_{\Gamma_\ell} \frac{\zeta^k \dd \zeta}{R(\zeta)},
\label{A-system}
\end{equation}
indexed by $k=0,1,2,\ldots,2n-1$. Note that the right-hand side of \eqref{A-system} and the coefficients of the unknowns $A_{\pm j}(x,t)$ are all purely imaginary for each $k$. Therefore, $A_{\pm j}(x,t)$ are real-valued. This is alarming at first; however, the coefficient matrix for this linear system is independent of $(x,t)$, and the right-hand side is bounded in $(x,t)$ (even though $\Omega_\ell(x,t) \in\mathbb{R}$ may, in principle, grow unboundedly). This structure ensures that $A_{\pm j}(x,t)\in\mathbb{R}$ are bounded in $(x,t)$. The integrals that make up linear system \eqref{A-system} are again computed numerically via Gauss--Chebyshev quadrature and by employing a suitable basis of polynomials. In particular, we now use 
\begin{equation*}
q_{k}(x)=\prod_{j=-n, j\neq 0,k}^{n}(x-\nu_j), \quad k=\pm1,\ldots,\pm n,
\end{equation*} 
where $\nu_j$ denotes the midpoint of the $j$th band ($(a_j,b_j)$ if $j>0$ or $(-b_{-j},-a_{-j})$ if $j<0$). 

\begin{figure}
\includegraphics[width=\textwidth]{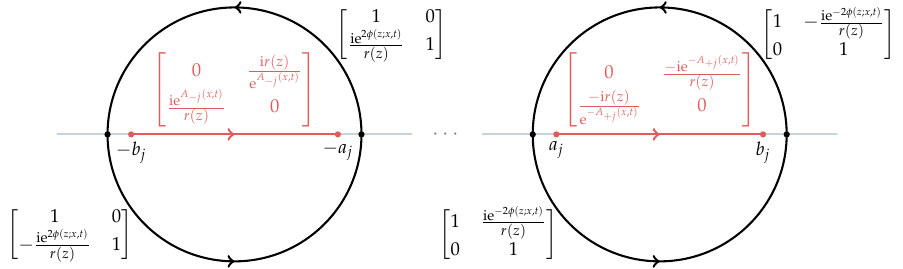}
\caption{Jump contours and jump matrices associated with $\mathbf{R}(z;x,t)$ near each interval. $\phi(z;x,t) = \varphi(z;x,t) + h(z;x,t)$.}
\label{f:R-jumps}
\end{figure}

We now make a global definition and introduce an exponent:
\begin{equation}
\mathbf{R}(z):= \mathbf{S}(z)\ee^{h(z;x,t)\sigma_3}\quad \text{and}\quad \phi(z;x,t):=h(z;x,t) + g(z;x,t) - \theta(z;x,t).
\end{equation}
Note that $\mathbf{R}(z) = \begin{bmatrix}1& 1 \end{bmatrix} + \mathrm{O}(z^{-1})$ as $z\to\infty$. Since the transformations $\mathbf{Y}(z)\mapsto\mathbf{S}(z)\mapsto\mathbf{R}(z)$ involve right-multiplications by diagonal unimodular matrices near $z=\infty$, it follows from \eqref{recovery} that
\begin{equation}
u(x,t) = \lim_{z\to\infty} 2z^2(r_1(z;x,t) r_2(z;x,t) - 1).
\end{equation}
If $z\in\Gamma_\ell$, then $h^+(z;x,t) - h^-(z;x,t) + \ii\Omega_{\ell}(x,t) = \log(\exp(-\ii \Omega_{\ell}(x,t)) +\ii\Omega_{\ell}(x,t) = \ii 2 k \pi$
for some $k\in\mathbb{Z}$, so $\mathbf{R}(z)$ has no jumps across the gaps $\Gamma_\ell$. The jump conditions satisfied by $\mathbf{R}(z)$ are described in Figure~\ref{f:R-jumps}.
This final RHP satisfied by $\mathbf{R}(z)$ is again treated numerically by using appropriate basis functions and an ansatz involving appropriate weighted Cauchy transforms, similar to \eqref{ansatz}. More specifically, we use Laurent polynomials for the components on the circular contours and appropriate Chebyshev polynomials and their weight functions for the components on $\Sigma_+ \cup \Sigma_-$, capturing the endpoint behavior of $\mathbf{R}(z)$, which is exactly the same as in $\eqref{S-endpt}$.
This approach closely follows the computational framework in \cite{BallewT24}.
\begin{remark}
This method could, in principle, be sped up significantly by considering $r$ with a common analytic extension to a region containing $[-b_n,b_n]$. In this case, the $2n$ circular jump contours can be replaced by a single large one. See \cite[Appendix A]{BallewT23}.
\end{remark}

Note that the gap in the $(x,t)$-plane for $t>0$ between the regions described in Section~\ref{s:quiet} and Section~\ref{s:unmodulated} disappears at $t=0$, allowing us to compute the soliton gas primitive potentials quickly and accurately for all $x\in\mathbb{R}$.

\section{Nonlinear superposition of a pure soliton gas with solitons} 

It is possible to insert a multi-soliton into a soliton gas by supplying suitable residue conditions for the RHP satisfied by $\mathbf{Y}(z)$ for a number of simple poles. In this case, we consider solving the following RHP:

\begin{rhp}[KdV soliton gas with solitons]
Let $(x,t)\in\mathbb{R}\times\mathbb{R}_{+}$ be fixed. Find a $1\times 2$ (row vector-valued) function $\mathbf{N}(\lambda)\equiv \mathbf{N}(\lambda;x,t)$ with the following properties:
\begin{itemize}
\item[]\textbf{Analyticity:} $\mathbf{N}(\lambda)$ is analytic for 
$\lambda\in\mathbb{C}\setminus\mathrm{cl}( \ii\Sigma_+ \cup \ii \Sigma_-)$, admitting continuous boundary values on $\ii\Sigma_+ \cup \ii \Sigma_-$, with the exception of simple poles at $\lambda=\pm\ii \kappa_j$, $\kappa_j>0$, $j=1,2,\ldots,N$, in the complement of the collection of intervals $\mathrm{cl}( \ii\Sigma_+ \cup \ii \Sigma_-)$.
\item[]\textbf{Jump conditions:} The boundary values $\mathbf{N}^+(\lambda)$ (resp.\@ $\mathbf{N}^-(\lambda)$) from left (resp.\@ from right) are related via
\begin{align}
\mathbf{N}^+(\lambda)&=\mathbf{N}^-(\lambda)\begin{bmatrix} 1 & 0 \\ -2\ii r_1(\lambda)\ee^{2\ii(x \lambda + 4 t \lambda^3)} & 1\end{bmatrix},\quad \lambda\in \ii(a_j,b_j),\\
\mathbf{N}^+(\lambda)&=\mathbf{N}^-(\lambda)\begin{bmatrix} 1 &  2\ii r_1(\lambda)\ee^{-2\ii(\lambda x+ 4 t \lambda^3)} \\ 0 & 1\end{bmatrix},\quad \lambda\in \ii(-b_j,-a_j), \quad j=1,2,\ldots,n.
\end{align}
\item[]\textbf{Residue conditions:} At each simple pole $\lambda=\pm\ii\kappa_j$, $j=1,2,\ldots,N$, $\mathbf{N}(z)$ satisfies
\begin{align}
\res_{\lambda=\ii \kappa_j} \mathbf{N}(z) = \lim_{\lambda\to\ii\kappa_j} \mathbf{N}(z) \begin{bmatrix} 0 & 0 \\ \ii \chi_j \ee^{2\ii (\ii \kappa_j x - 4\ii\kappa_j^3 t)} & 0 \end{bmatrix},\\
\res_{\lambda=-\ii \kappa_j} \mathbf{N}(z) = \lim_{\lambda\to-\ii\kappa_j} \mathbf{N}(z) \begin{bmatrix} 0 & -\ii \chi_j \ee^{-2\ii (\ii \kappa_j x - 4\ii\kappa_j^3 t)}\\ 0 & 0 \end{bmatrix},
\end{align}
with norming constants $\chi_j\in\mathbb{R_+}\setminus\{0\}$.
\item[]\textbf{Normalization:} $\mathbf{N}(\lambda)=\begin{bmatrix} 1 & 1\end{bmatrix} + \mathrm{O}(\lambda^{-1})$ as $\lambda\to\infty$.
\item[]\textbf{Symmetry condition:} $\mathbf{N}(-\lambda)=\mathbf{N}(\lambda) \begin{bmatrix} 0 & 1 \\ 1 & 0 \end{bmatrix}$.
\end{itemize}
\label{rhp:gas-soliton}
\end{rhp}

\begin{figure}
 \includegraphics[width=\linewidth]{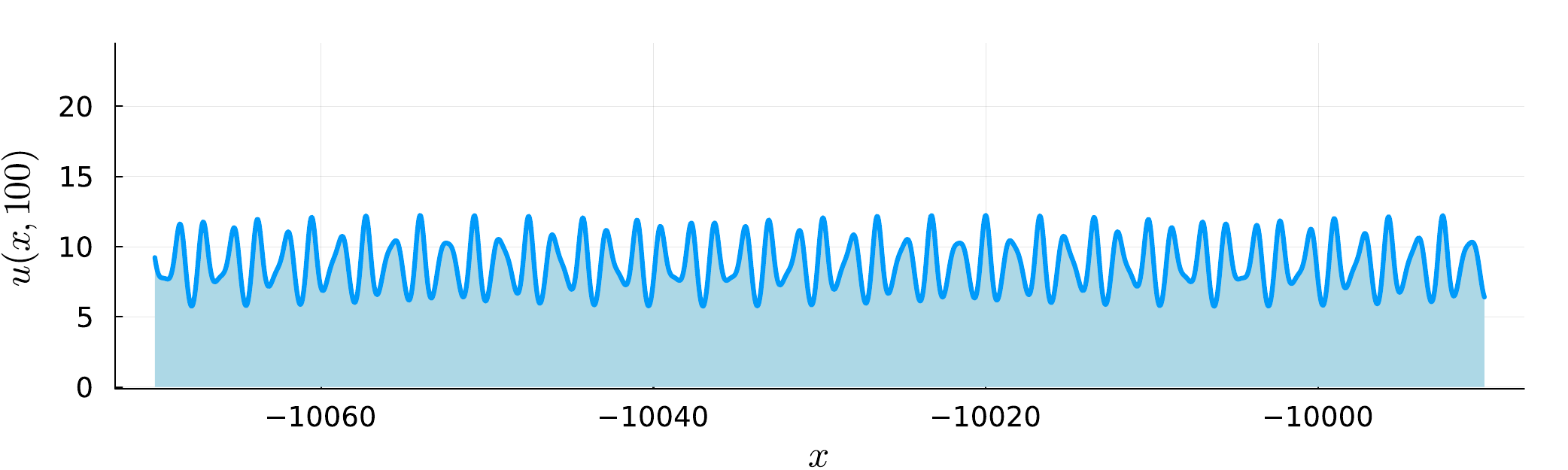}\\[2em]
 \includegraphics[width=\linewidth]{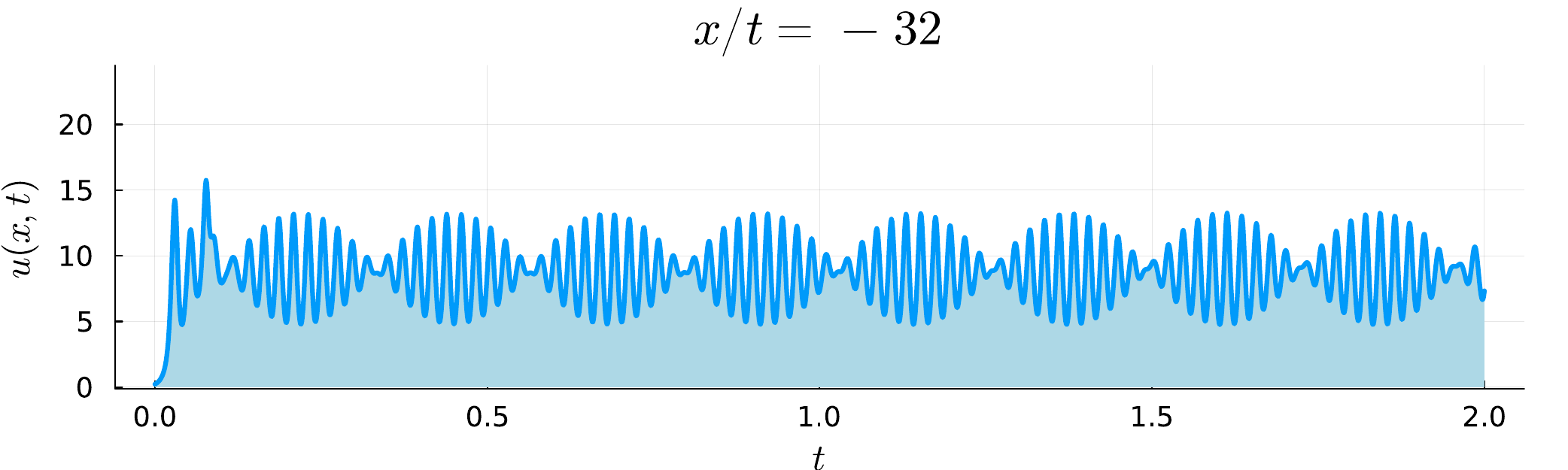}
 \caption{A KdV soliton gas with $r_1(\lambda)$ supported on two pairs of bands. In the notation of \eqref{r1-practice}, $I_1=(1,2)$, $I_2=(2.5,3)$ with $f_1(z)=100$, $f_2(z)=1$ and $\alpha_{j}=\beta_{j}=\frac{1}{2}$, $j=1,2$. The three solitons superimposed are associated with $\kappa_1=0.8$, $\kappa_2=2.25$, and $\kappa_3=3.5$ and the norming constants $\chi_1=10^6$, $\chi_2=10^5$, $\chi_3=10^{-12}$.
Bottom panel: The same solution plotted along the ray $x/t = -32$.}
    \label{fig:2-band-2-poles}
 \end{figure}

\begin{figure}
\includegraphics[width=\linewidth]{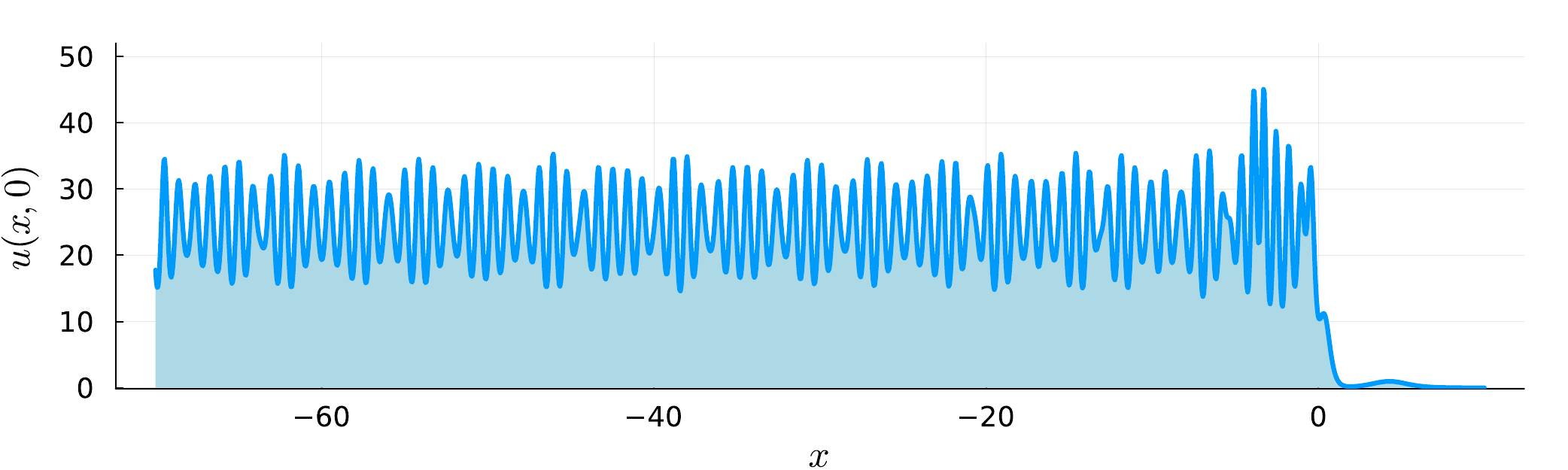}\\
\includegraphics[width=\linewidth]{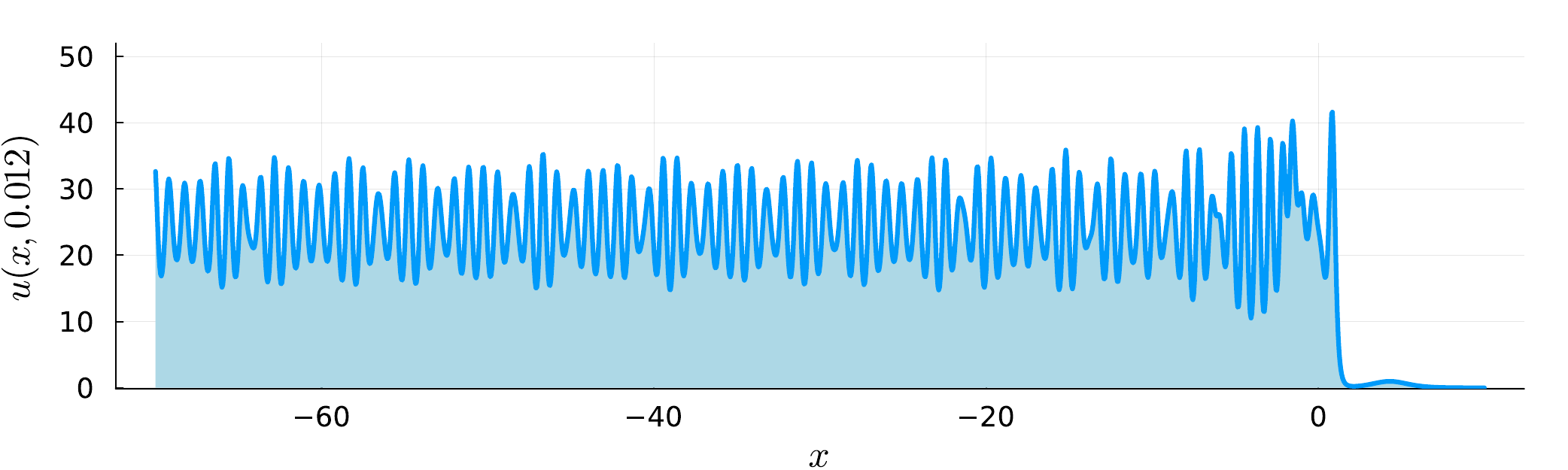}
 \caption{A KdV soliton gas with $r_1(\lambda)$ supported on five pairs of bands, nonlinearly superimposed with five solitons. In the notation of \eqref{r1-practice}, $I_1=(0.25,0.5)$, $I_2=(0.8,1.2)$, $I_3=(1.5,2)$, $I_4=(2.5,3)$, and $I_5=(4,5)$ with $f_1(z)=(z-0.375)^2+1$, $f_2(z)=(z-1)^4+1$, $f_3(z)=(z-1.75)^6+1$, $f_4(z)=\exp(z-2.75)+1$, $f_5(z)=\exp(-(z-4.5)^2)+1$ and $\alpha_1=\beta_1=-\frac{1}{2}$, $\alpha_2=\beta_2=\frac{1}{2}$, $\alpha_3=\frac{1}{2}$, $\beta_3=-\frac{1}{2}$, $\alpha_{4}=-\frac{1}{2}$, $\beta_4=\frac{1}{2}$, $\alpha_5=\beta_5=-\frac{1}{2}$. The five solitons are associated with the eigenvalue parameters $\kappa_1=0.1$, $\kappa_2=0.7$, $\kappa_3=2.25$, $\kappa_4=3.5$, $\kappa_5=5.5$ and the norming constants $\chi_1=10^5$, $\chi_2=1000$, $\chi_3=100$, $\chi_4=10$, $\chi_5=10^{-6}$.}
 \label{fig:fvar}
 \end{figure}

As with \rhref{rhp:gas}, we rotate the $\lambda$-plane and consider the related function $\mathbf{Y}(z):=\mathbf{N}(\ii z)$. The jump contours and pole locations $z=\pm\kappa_j$ for $\mathbf{Y}(z)$ are then on the real axis. 
The solution $u(x,t)$ of the KdV equation \eqref{KdV} is obtained from \eqref{recovery}. In Figure~\ref{fig:2-band-2-poles}, we present the computed large-time evolution of a soliton gas supported on two pairs of bands, nonlinearly superimposed with three solitons. In the notation of \eqref{r1-practice}, we choose $f_1(z)$ to be much larger than $f_2(z)$.  We also note here that our method is by no means limited to constant functions $f_j(z)$ in \eqref{r1-practice} (see Figure~\ref{fig:fvar}).

\begin{figure}
 \includegraphics[width=\linewidth]{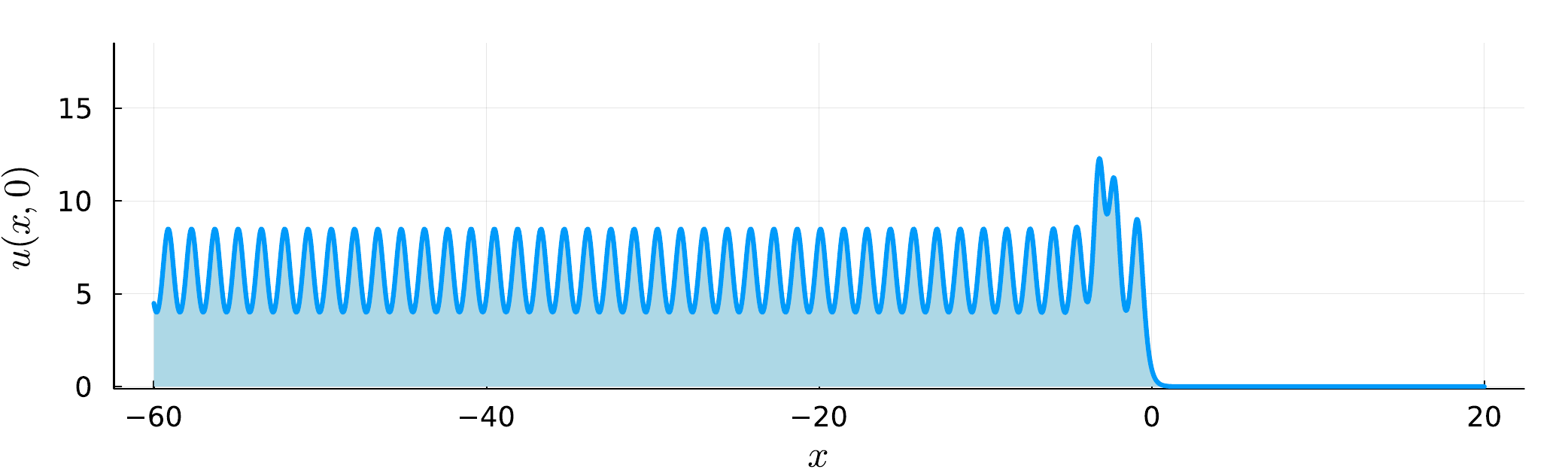}\\[2em]
 \includegraphics[width=\linewidth]{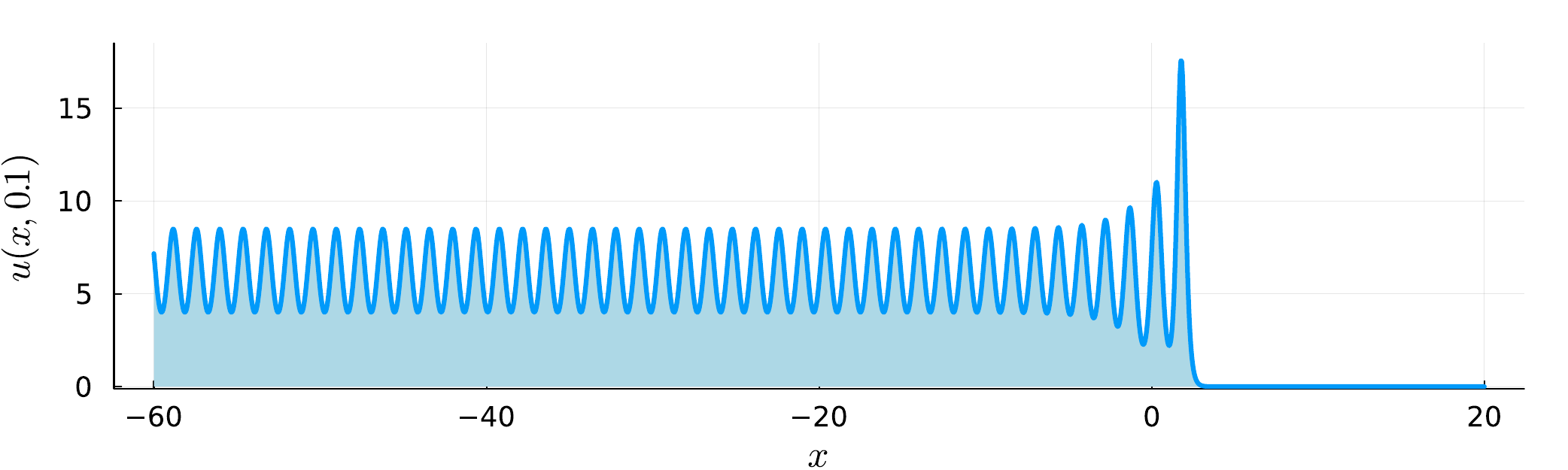}\\[2em]
 \includegraphics[width=\linewidth]{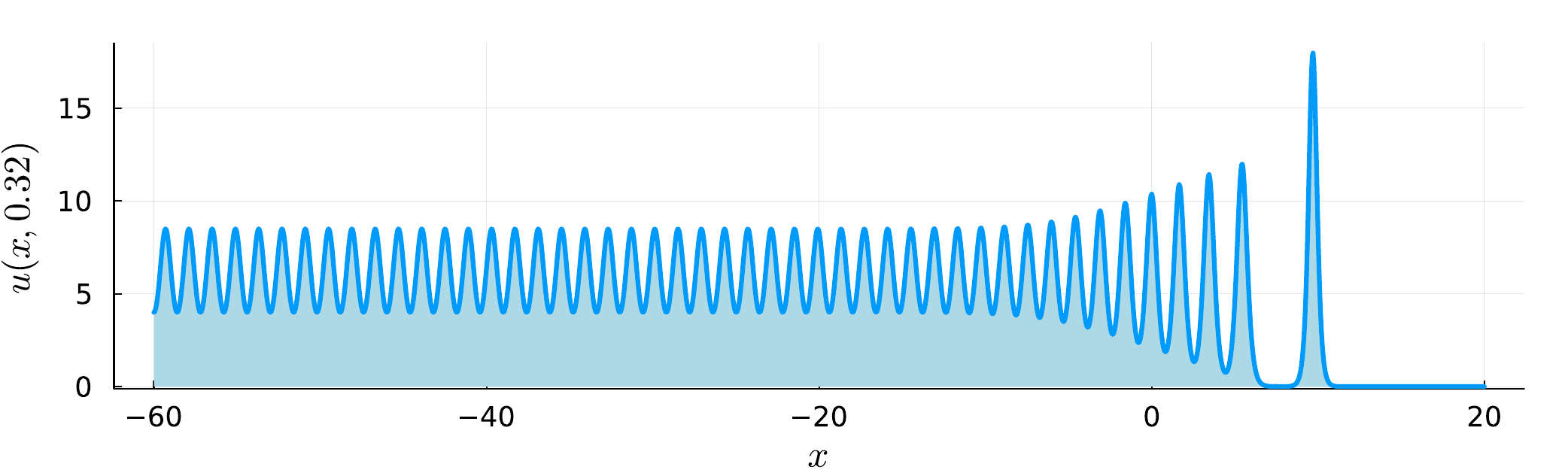}
 \caption{A KdV soliton gas with $r_1(\lambda)$ supported on one pair of bands. In the notation of \eqref{r1-practice}, $I_1=(1.5,2.5)$ with $f_1(z)=1$ and $\alpha_{1}=\beta_{1}=\frac{1}{2}$. The superimposed soliton is associated with $\kappa_1=3$ and the norming constant $\chi_1=10^{-4}$.}
    \label{fig:tall-sol}
 \end{figure}

 \begin{figure}
 \includegraphics[width=\linewidth]{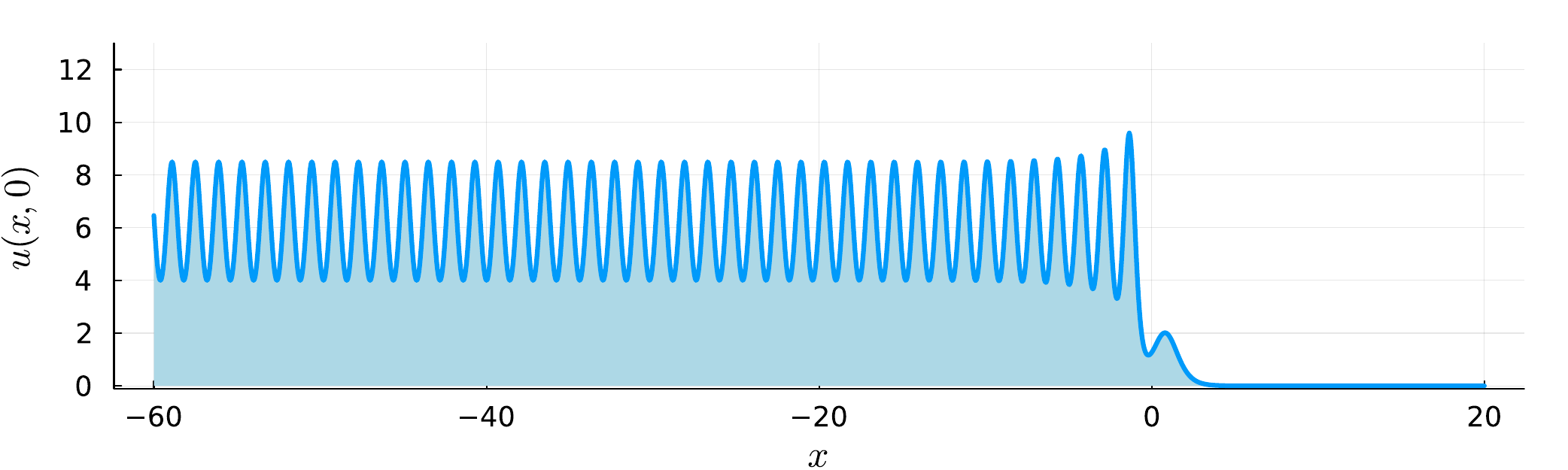}\\[2em]
 \includegraphics[width=\linewidth]{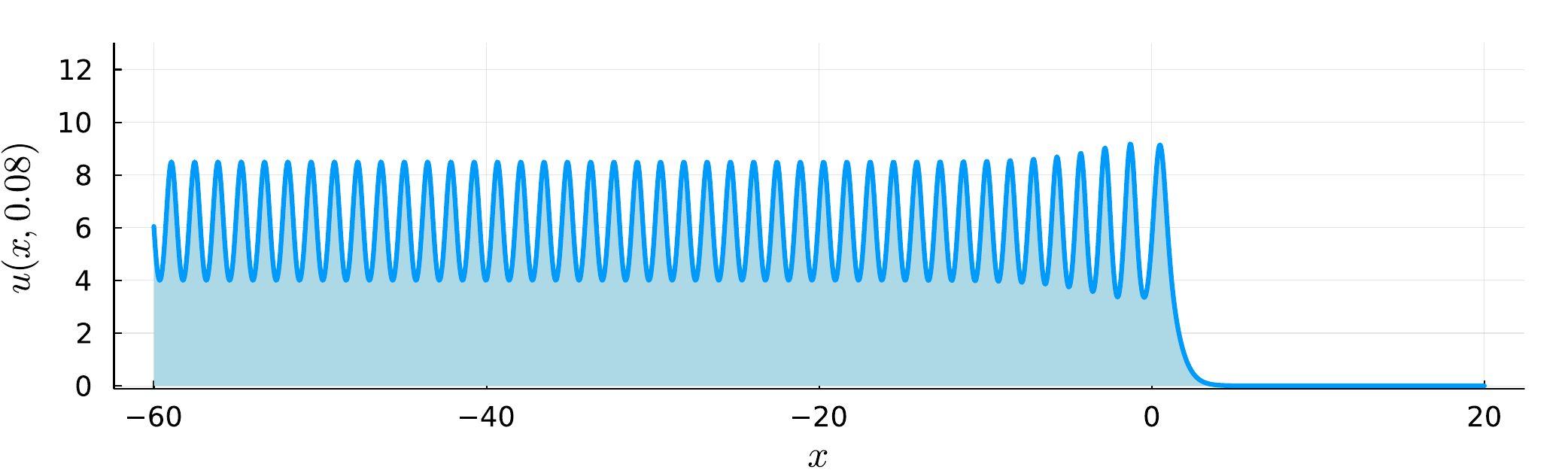}\\[2em]
 \includegraphics[width=\linewidth]{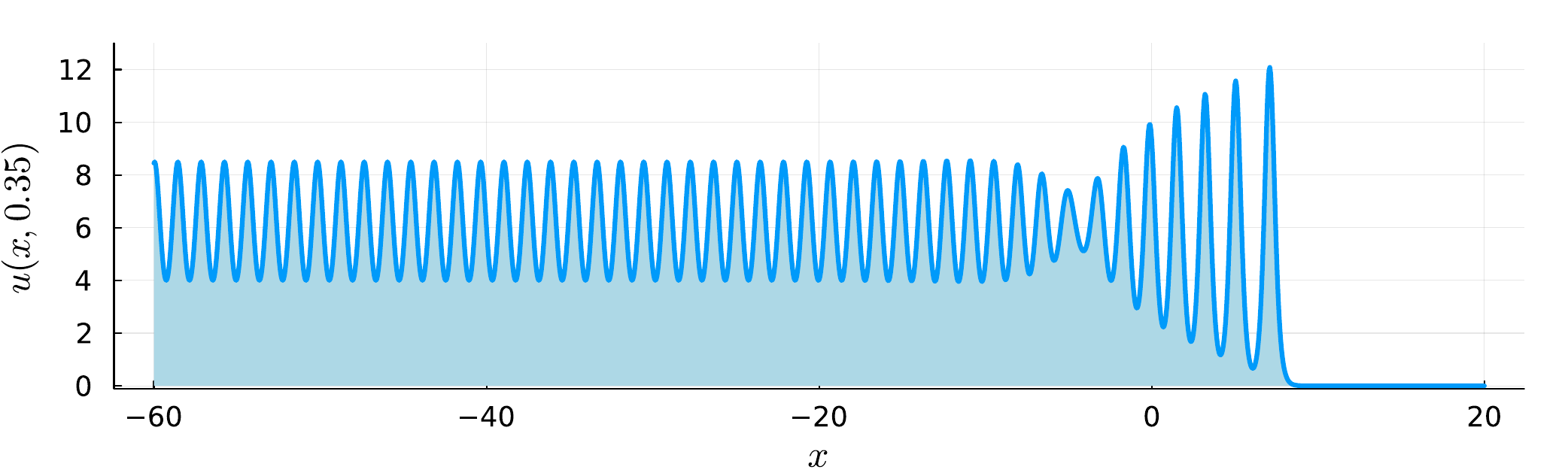}
 \caption{A KdV soliton gas with $r_1(\lambda)$ supported on one pair of bands. In the notation of \eqref{r1-practice}, $I_1=(1.5,2.5)$ with $f_1(z)=1$ and $\alpha_{1}=\beta_{1}=\frac{1}{2}$. The superimposed soliton is associated with $\kappa_1=1$ and the norming constant $\chi_1=10$.}
    \label{fig:trapped-sol}
 \end{figure}

\subsection{The solution procedure}
An overview of our procedure is as follows:  We first ignore the poles in \rhref{rhp:gas-soliton} (in the rotated plane of $z=-\ii \lambda$) and compute the matrix solution\footnote{A matrix solution ceases to exist at a set of isolated points $(x,t)$, but we ignore this behavior for our numerical purposes just as in \cite{BilmanT17} and \cite{TrogdonOD12}.} $\mathbf{P}(z)$, normalized so that $\mathbf{P}(z)\to \mathbb{I}$ as $z\to\infty$, and use $\mathbf{P}(z)$ as a global parametrix for the RHP with poles, reducing it to a discrete finite-dimensional linear algebra problem. We solve the discrete problem, computing the solution $\vec S(z)$, and superpose its solution via $\vec S(z) \mathbf{P}(z)$ to compute the nonlinear superposition of solitons with soliton gasses. We now describe the steps in this procedure in more detail.

To handle potential exponential growth in the residue conditions, we flip their triangularities as needed via the transformation
\begin{equation*}
\hat{\vec Y}(z)=\vec Y(z)v(z)^{\sigma_3},\quad v(z)=\prod_{j\in K}\frac{z-\kappa_j}{z+\kappa_j},
\end{equation*}
where $K$ is an index set corresponding to ``large'' residue conditions. In particular, for some constant $c>0$, we define 
\begin{equation*}
K = K(x,t) =\left\{j:\left|\chi_j\ee^{8\kappa_j^3t-2\kappa_j x}\right|>c\right\},
\end{equation*}
in the quiescent region and
\begin{equation*}
K=\left\{j:\left|\chi_j\ee^{8\kappa_j^3t-2\kappa_j x+2g(z)+2h(z)}\right|>c\right\},
\end{equation*}
in the unmodulated region. Our implementation takes $c=10$ as the default value.

This transformation has the effect of modifying the residue conditions to 
\begin{align*}
&\res_{z=\kappa_j}\hat{\vec Y}(z)=\lim_{z\to\kappa_j}\hat{\vec Y}(z)\begin{bmatrix}0 & 0 \\
    \chi_j\ee^{8\kappa_j^3t-2\kappa_j x}v^2(\kappa_j)& 0
\end{bmatrix},\quad j\notin K,\\
&\res_{z=\kappa_j}\hat{\vec Y}(z)=\lim_{z\to\kappa_j}\hat{\vec Y}(z)\begin{bmatrix}0 & \frac{\ee^{-8\kappa_j^3t+2\kappa_j x}}{\chi_jv'(\kappa_j)^2} \\
    0& 0
\end{bmatrix},\quad j\in K.
\\
\end{align*}
The residue conditions on the negative real axis follow from the symmetry condition.

In the unmodulated region, the disjoint disks $D_j$ are assumed to be sufficiently small so as not to contain the poles $\kappa_j$. The steepest descent deformations then result in the residue conditions
\begin{align*}
&\res_{z=\kappa_j}\vec R(z)=\lim_{z\to\kappa_j}\vec R(z)\begin{bmatrix}0 & 0 \\
    \chi_j\ee^{8\kappa_j^3t-2\kappa_j x+2g(\kappa_j)+2h(\kappa_j)}v^2(\kappa_j)& 0
\end{bmatrix},\quad j\notin K,\\
&\res_{z=\kappa_j}\vec R(z)=\lim_{z\to\kappa_j}\vec R(z)\begin{bmatrix}0 & \frac{\ee^{-8\kappa_j^3t+2\kappa_j x-2g(\kappa_j)-2h(\kappa_j)}}{\chi_jv'(\kappa_j)^2} \\
    0& 0
\end{bmatrix},\quad j\in K.
\end{align*}

Once the matrix solution $\vec P$ is obtained, the residue conditions that $\vec S(z)=\vec R(z)\vec P(z)^{-1}$ satisfies are computed through the following formula: Given a residue condition 
\begin{equation*}
\res_{z=\pm\kappa_j}\vec R(z)=\lim_{z\to \pm\kappa_j}\vec R(z)\vec \Xi_{\pm j},
\end{equation*}
of $\vec R$, the corresponding residue condition of $\vec S$ is given by
\begin{equation*}
\res_{z=\pm\kappa_j}\vec S(z)=\lim_{z\to \pm\kappa_j}\vec S(z)\vec P(\pm\kappa_j)\Xi_{\pm j}\vec P(\pm\kappa_j)^{-1}\left(\mathbb{I}-\vec P'(\pm\kappa_j)\Xi_{\pm j}\vec P(\pm\kappa_j)^{-1}\right)^{-1},
\end{equation*}
where $\vec P'(z)$ denotes the componentwise derivative of $\vec P(z)$. We compute $\vec P'(\pm\kappa_j)$ by numerically expanding $\vec P(z)$ in a Laurent series in a small circle centered at $\pm\kappa_j$.

\section{Examples, demonstration of convergence, and performance}
\subsection{Examples}
In this subsection, we enumerate the examples found throughout this paper.
\subsubsection{5 pairs of bands with 5 solitons} In Figures~\ref{fig:soliton-gas-wide-1}~and~\ref{fig:fvar}, we plot a soliton gas corresponding to $r_1(\lambda)$ supported on five pairs of bands, nonlinearly superimposed with five solitons. In the notation of \eqref{r1-practice}, Figure~\ref{fig:fvar} considers variable $f_j$ and various endpoint behaviors $\alpha_j,\beta_j$.
\subsubsection{5 pairs of bands} In Figure~\ref{fig:soliton-gas-tail}, we plot a pure soliton gas corresponding to $r_1(\lambda)$ supported on five pairs of bands, including for large negative $x$ at $t=100$.
\subsubsection{1 pair of bands with 2 solitons} In Figure~\ref{fig:wedge}, we consider a soliton gas corresponding to $r_1(\lambda)$ supported on one pair of bands, nonlinearly superimposed with two solitons. We include a density plot of this solution in the $(x,t)$-plane outside of the wedge-shaped region where our method begins to break down.
\subsubsection{2 pairs of bands} In Figure~\ref{fig:2-band}, we consider a pure soliton gas corresponding to $r_1(\lambda)$ supported on two pairs of bands. We plot its tail at $t=100$ and its behavior along the ray $x/t=-32$.
\subsubsection{2 pairs of bands with 3 solitons} In Figure~\ref{fig:2-band-2-poles}, we include the same plots as Figure~\ref{fig:2-band}, but with the solution nonlinearly superimposed with three solitons.
\subsubsection{1 pair of bands with a taller soliton} In Figure~\ref{fig:tall-sol}, we plot the time evolution of a soliton gas corresponding to $r_1(\lambda)$ supported on one pair of bands, nonlinearly superimposed with one taller soliton, showing the soliton escape the soliton gas.
\subsubsection{1 pair of bands with a shorter soliton} In Figure~\ref{fig:trapped-sol}, we plot the time evolution of a soliton gas corresponding to $r_1(\lambda)$ supported on one pair of bands, nonlinearly superimposed with one shorter soliton, showing the soliton becoming trapped within the soliton gas.
\subsection{Convergence}
\begin{figure}
\includegraphics[width=\linewidth]{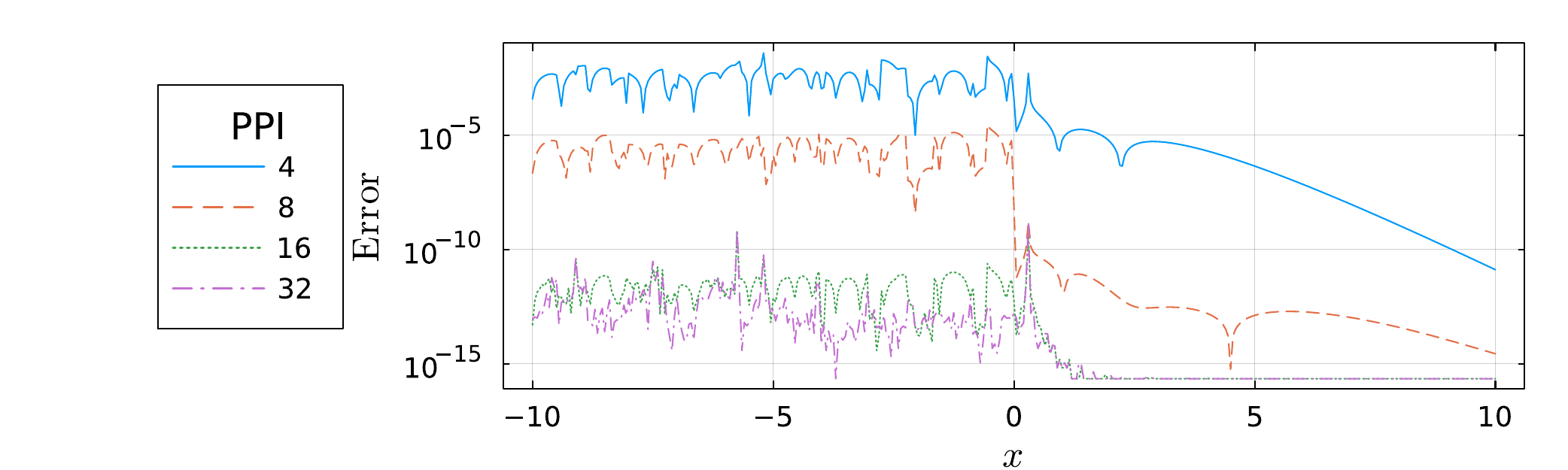}
\caption{Pointwise errors of various numbers for collocation points. PPI (Points Per Interval) denotes the number of collocation points used per interval $I_j$, and 10 times that number of points are used on each circle. 50 PPI is used as the exact solution in comparisons.}
\label{fig:errors}
\end{figure}

In Figure~\ref{fig:errors}, we demonstrate the convergence of our computational method as we increase the number of collocation points used on each component of the jump contour associated with the RHP that is treated numerically. 
We plot the absolute pointwise error made in computing a soliton gas potential
\begin{equation}
E_\mathrm{p}(x):=|u_{\mathrm{p}}(x,0) - u_{\mathrm{true}}(x,0)|,
\end{equation}
where $\mathrm{p}$ refers to the number of collocation points used on each interval, denoted by ``PPI" in Figure~\ref{fig:errors}. Here, $u_{\mathrm{true}}(x,0)$ denotes the solution computed with a large value of $\mathrm{p}$.
The soliton gas potential computed for Figure~\ref{fig:errors} is associated with $r_1(\lambda)$ supported on two pairs of bands (in the notation of \eqref{r1-practice}) $I_1=(1.2,2)$, $I_2=(2.5,3)$ with $f_1(z)=100$ and $f_2(z)=1$ and $\alpha_{j}=\beta_{j}=\frac{1}{2}$, $j=1,2$. This soliton gas potential is also nonlinearly superimposed with two solitons associated with the eigenvalue parameters $\kappa_1=1$, $\kappa_2=4$ and the norming constants $\chi_1=10$, $\chi_2=10^{-10}$. While we observe that our method is quite accurate, some precision is lost due to the large condition number of the collocation matrix (empirically on the order of $10^7$). Future work will aim to improve the conditioning of this linear system.

Finally, in all of the plots presented in this paper, $20$ collocation points are used on each interval and $120$ points are used on each circle in the jump contour for the RHP treated numerically.

\subsection{Notes on performance}
Perhaps the most attractive feature of our method is that it is trivially parallelizable. Since $x$ and $t$ are only parameters in the RHP, all computations in this paper can, in principle, be done in the time it takes to solve a single RHP, i.e., a pointwise evaluation of $u(x,t)$. With this perspective in mind, we give rough pointwise evaluation runtimes on a standard laptop for the solution plots above: Figures~\ref{fig:soliton-gas-wide-1}~and~\ref{fig:fvar} require roughly 0.7 seconds in the unmodulated region and 0.08 seconds in the quiescent region. Figure~\ref{fig:soliton-gas-tail} requires roughly 0.4 seconds in the unmodulated region and 2 ms in the quiescent region. Figure~\ref{fig:wedge} requires roughly 0.03 seconds in the unmodulated region and 5 ms in the quiescent region. Figure~\ref{fig:2-band} requires roughly 0.02 seconds in the unmodulated region and 0.4 ms in the quiescent region. Figure~\ref{fig:2-band-2-poles} requires roughly 0.1 seconds in the unmodulated region and 0.02 seconds in the quiescent region. Figures~\ref{fig:tall-sol}~and~\ref{fig:trapped-sol} require roughly 0.02 seconds in the unmodulated region and 3 ms in the quiescent region.

\end{document}